\documentclass[journal,10pt]{IEEEtran}
\usepackage{amsmath,amsfonts}
\usepackage{array}
\usepackage{xcolor} 
\usepackage{textcomp}
\usepackage{stfloats}
\usepackage{url}
\usepackage{comment}
\usepackage[colorlinks=true, linkcolor=blue, citecolor=green, urlcolor=cyan]{hyperref}
\usepackage{amssymb}
\usepackage{booktabs}
\usepackage{multirow}
\usepackage{makecell} 
\usepackage{verbatim}
\usepackage{graphicx}
\usepackage{stmaryrd}
\usepackage{algorithm}

\usepackage{algpseudocode} 
\usepackage{cite}
\usepackage[caption=false,font=normalsize,labelfont=sf,textfont=sf]{subfig}

\begin{document}
	\title{Achieving Constant-Envelope Waveform in CP-OFDMA Framework}
	
	\author{Yiming~Zhu,~\IEEEmembership{Graduate Student~Member,~IEEE}, 
		Zhuhong~Zhu,  
		Xiaodong~Xu, 
		Hongwei~Hou,~\IEEEmembership{Graduate Student~Member,~IEEE}, 
		Wenjin~Wang,~\IEEEmembership{Member,~IEEE}, 
		and 
		Rui Ding%
		\thanks{Manuscript received xxx; revised xxx; accepted xxx.  
		\textit{(Yiming Zhu and Zhuhong Zhu contributed equally to this work.) (Corresponding author: Wenjin Wang.)}}
		\thanks{Yiming Zhu, Zhuhong Zhu, Hongwei Hou, and Wenjin Wang are with the National Mobile Communications Research Laboratory, Southeast University, Nanjing 210096, China, and also with Purple Mountain Laboratories, Nanjing 211100, China (e-mail: ymzhu@seu.edu.cn; 
		220230836@seu.edu.cn; 
		hongweihou@seu.edu.cn;  
		wangwj@seu.edu.cn).} 
		\thanks{Xiaodong Xu is with China Mobile Research Institute, Beijing 100032, China (e-mail: xuxiaodong@chinamobile.com).}
		\thanks{Rui Ding is with China Satellite Network Group Company Ltd., Beijing 100029, China (e-mail: greatdn@qq.com).}
	}
	
\maketitle

\begin{abstract}
	Orthogonal frequency division multiplexing (OFDM) is widely adopted in modern wireless communication systems, but its power efficiency is limited by high envelope fluctuations.	 
	Although various high power-efficiency waveforms have been proposed, most are incompatible with the cyclic-prefixed orthogonal frequency division multiple access (CP-OFDMA) framework and remain ineffective in multi-user downlink transmissions. 
	To address this issue, we propose a constant-envelope (CE) waveform design, which enables low-complexity transceiver architectures while maintaining full compatibility with the prevailing CP-OFDMA framework. 
	Specifically, we start from a general CE FDMA signal model and develop a CP-OFDMA-compatible waveform implementation structure, followed by the design of an optimized  CE-constrained pulse-shaping filter to suppress out-of-band emissions. 
	To tackle channel estimation challenge under non-flat frequency-domain pilots induced by CE modulation, we optimize the time-domain binary pilot sequence to achieve frequency-domain CE properties, and then propose a multi-stage method combining delay-domain denoising with power delay profile estimation to facilitate reduced-dimension linear minimum mean square error (LMMSE) estimation. 
	Subsequently, we design a low-complexity maximum ratio combining-aided LMMSE equalizer by exploiting the periodicity and conjugate symmetry of the CE received signals. 
	To mitigate the downlink peak-to-average power ratio increase caused by FDMA, we further develop a multi-user downlink CE transmission scheme including multiple access mechanism, downlink control information design, and corresponding system-level implementation, which ensures compatibility with the New Radio standard. 
	Numerical results demonstrate that the proposed scheme achieves bit error rate performance close to the ideal case while significantly reducing transceiver complexity compared to existing CE waveform~solutions. 
\end{abstract}


\section{Introduction}
\IEEEPARstart{O}{rthogonal} frequency division multiplexing (OFDM), which has long been entrenched in wireless standards due to its robustness against frequency-selective fading, low implementation complexity, and suitability for multiple-antenna transmission, is anticipated to persist as a fundamental modulation technique in future cellular networks \cite{BroadbandMIMO, OFDMorW,zhu2022ofdm}. 
However, the OFDM-based waveform suffers from an inherently high peak-to-average power ratio (PAPR), which forces power amplifiers (PAs) to operate with significant output back-off to maintain linearity, thereby drastically reducing their power efficiency and potentially introducing nonlinear distortion when the PA operates near saturation \cite{yang2019sixg}. 
The resulting decline in PA efficiency and signal fidelity poses critical challenges for several power-constrained sixth-generation (6G) wireless scenarios, notably the Internet of Things (IoT) and non-terrestrial networks (NTNs). 
Specifically, IoT devices are typically subject to limited battery capacity and minimal radio-frequency front-end complexity, which in turn necessitates power-efficient waveform designs to sustain long-term operation \cite{yang2019sixg,SingleCarrieIn}.
Similarly, NTNs, including satellite and aerial platforms, also demand high-efficiency transmissions due to limited onboard power and extended propagation distances~\cite{wang2025toward,wu2023energy,NearOptimalTiming,wu2025distributed,MassiveMIMOTra}. These limitations motivate the development of low-PAPR waveform solutions that ensure high PA efficiency and mitigate nonlinear degradation, which are essential for energy-efficient transmission in future 6G networks. 

Improving the power efficiency of OFDM waveform has long been a key research objective \cite{OFDMInsp}. One of the most widely adopted approach is discrete Fourier transform-spread-OFDM (DFT-s-OFDM)-based single-carrier frequency division multiple access (SC-FDMA), which reduces the high PAPR of OFDM signals while retaining compatibility with  legacy orthogonal frequency division multiple access (OFDMA) architectures \cite{lecarrie}. 
To further enhance power efficiency, various constant-envelope (CE) waveform schemes—such as CE-OFDM, CE single-carrier OFDM (CE-SC-OFDM), and CE-OFDM with index modulation (CE-OFDM-IM)—have been proposed, all of which generate CE waveforms by applying nonlinear phase modulation to the OFDM signals~\cite{ConstantEnvelopeOFD,nstantenvel,chen2024constant,wang2025ceofdm}. 
While these schemes effectively suppress envelope fluctuations, they also introduce notable drawbacks. In particular, the use of phase modulation increases sensitivity to phase noise and necessitates complicated  phase-demodulation algorithms at the receiver \cite{onstantEnvelopeMultica}. 
In parallel, the authors in \cite{LocalizedSC} proposed another CE waveform scheme, where a minimum shift keying (MSK) modulator is applied after the Gaussian filter in the SC-FDMA transmitter to achieve CE transmission, but still facing the high demodulation complexity caused by nonlinear modulation. 
Moreover, efforts have been devoted to reduce PAPR while preserving low-complexity equalization. A promising direction is the incorporation of offset quadrature amplitude modulation (OQAM) in OFDM-based systems \cite{Gao2011Cyclic, Wang2013CP}. 
To jointly exploit the low PAPR characteristics of OQAM-modulated single-carrier signals and the efficient frequency-domain equalization enabled by the CP-OFDM framework, a cyclic prefixed version of the OQAM based OFDM (CP-OQAM-OFDM) has been developed and applied to SC-FDMA transmission \cite{Gao2011Cyclic, Wang2013CP}. 
	
From the preceding state-of-the-art overview, existing  CE waveform designs typically suffer from high receiver complexity, which restricts their deployment in cost- and power-constrained scenarios. 
While CP-OQAM-OFDM reduces PAPR and supports low-complexity equalization, it does not guarantee a strictly CE signals, thereby degrading PA efficiency in highly nonlinear radio frequency front-ends and limiting power savings. 
Furthermore, previous research has predominantly focused on multi-user uplink transmission, while relatively few efforts addressing CE waveform designs for multi-user downlink transmission. 
	
	Motivated by the preceding discussions, this paper develops a CE waveform design within the cyclic-prefixed OFDMA (CP-OFDMA) framework, targeting low-complexity transceiver processing and support for multi-user downlink transmission. The main contributions are summarized as follows: 
	\begin{itemize}
		\item 
		\textbf{CP-OFDMA-Compatible CE Waveform Design}: 
		To minimize the PAPR while maintaining low-complexity transmitter structure, we start from a CE FDMA waveform and develop a CE waveform implementation structure that is compatible with the legacy CP-OFDMA framework with minimal modifications. In order to jointly minimize the out-of-band energy and preserve the CE property, we further design the CE and near CE (NCE) pulse-shaping filters that effectively mitigate spectral leakage-induced multi-user access interference (MAI). 
		\item \textbf{Receiver Design For CE-CP-OFDMA}:  
		Based on the proposed CE waveform, we propose a low-complexity receiver architecture compatible with CP-OFDMA. We first establish the received signal model over frequency-selective fading channels, where the CE constraint results in non-flat frequency-domain signals. 
		To address the resulting challenge in channel estimation, we optimize the time-domain binary pilot sequence with the objective of frequency-domain CE, and then develop a multi-stage channel estimation method, which combines delay-domain denoising with power delay profile (PDP) estimation to facilitate linear minimum mean square error (LMMSE) channel estimation.  
		Moreover, we exploit the periodicity and conjugate symmetry of the CE received signal to perform maximum ratio combining (MRC), followed by a low-complexity frequency-domain LMMSE to mitigate inter-symbol interference (ISI).
		\item \textbf{CE Downlink Multiple Access Transmission}: 
		To address the PAPR increase caused by time-domain superposition in multi-user FDMA downlink transmission, 
		we propose two novel CE multiple access mechanisms: symbol-level multiple access (SLMA) and bit-level multiple access (BLMA). Furthermore, the corresponding two CE downlink control information (CE-DCI) formats are designed to enable these multiple access mechanisms. 
		To ensure compatibility with the NR system, we design a CE physical downlink control channel (CE-PDCCH) for CE-DCI transmission and propose a system-level downlink transmission scheme that incorporates the proposed CE multiple access mechanisms into the NR system with minimal modifications. 
	\end{itemize}
	
	The remainder of this paper is outlined as follows. Section \ref{Section_3} presents the proposed CE CP-OFDM (CE-CP-OFDM) signal model and provides the compatible implementation structure of the proposed schemes along with the optimized design of CE pulse-shaping filters. Section \ref{Section_4} proposes a efficient receiver architecture specifically tailored for the CE waveform. Section \ref{Section_5} develops a multi-user downlink CE transmission scheme compatible with the NR standard. Simulations and analyses of the proposed schemes are performed in Section~\ref{Section_6}. Finally, we conclude our work in Section \ref{Section_Conclusion6}.
	
	{\it Notations}: $\mathbf{I}_{M}$ and $\mathbf{J}_{M}$ are the $M \times M$ identity and reverse identity matrices, respectively. 
	$\mathbf{e}_{M, k}$ is the $k$-th column vector of the identity matrix $\mathbf{I}_{M}$. 
	$\mathbf{0}_{M \times N}$ and $\mathbf{1}_{M \times N}$ denote all-zero and all-one matrices of size $M \times N$, respectively. 
	$\mathbb{C}$ and $\mathbb{R}$ represent complex number and real number fields, respectively. 
	The superscripts $(\cdot)^*$, $(\cdot)^T$, $(\cdot)^H$, $(\cdot)^{-1}$ and $(\cdot)^{\dagger}$ denote conjugate, transpose, conjugate transpose, inverse, and pseudo-inverse, respectively. 
	$||\cdot||_{\mathrm{F}}$ and $||\cdot||_2$ denote the Frobenius norm and the $l_2$ norm, respectively. 
	$\left|\mathbf{d}\right|^{\odot 2}$ denotes the element-wise squared modulus of $\mathbf{d}$. 
	$\otimes$, $\odot$, and $ \circ $  denote the Kronecker, Hadamard, and Khatri-Rao products, respectively. 
	$\mathsf{diag}\{\mathbf{d}\}$ denotes a diagonal matrix with $\mathbf{d}$ on the main diagonal. 
	$ \mathsf{E}(\cdot) $ and $ \mathsf{Cov}(\cdot) $ denotes the expectation and covariance operators, respectively. 
	$\mathsf{r}(\mathbf{A})$ and $\mathsf{tr}(\mathbf{A})$ denote the rank and trace of matrix $\mathbf{A}$, respectively. 
	$[\mathbf{A}]_{m,n}$, $[\mathbf{A}]_{m:n,:}$ and $[\mathbf{A}]_{:,m:n}$ denote the $(m,n)$-th entry, the submatrix from rows $m$ to $n$, and the submatrix from columns $m$ to $n$ of $\mathbf{A}$, respectively. 
	$d(m)$ denotes the $m$-th entry of the vector $\mathbf{d}$.  
	$ |\cdot| $, $\lceil \cdot \rceil$, and $\lfloor \cdot \rfloor$ stand for the absolute, ceiling, and floor operators, respectively. 
	$\langle m \rangle_M$ denotes $m$ modulo $M$. 
	 $\mathbf{W}_{N} $ denotes the normalized $N$-point DFT matrix with the $ (m,n) $-th element $ \frac{1}{{\sqrt {N} }}{e^{ - j2\pi mn/N}}$.

	\section{Constant-Envelope Waveform in CP-OFDMA Framework} \label{Section_3}
	\subsection{Constant-Envelope SC-FDMA}
	We consider an uplink FDMA system with $K$ user equipments (UEs), where $K$ parallel data streams are encoded, interleaved, mapped separately, and pass through their respective pulse-shaping filters. To achieve better PAPR performance, each UE employs an  OQAM modulated single-carrier waveform. The resulting continuous-time baseband transmitted signal of the $k$-th UE is expressed as \cite{Wang2013CP} 
	\begin{equation}
		{{\chi}_{\mathrm{C},k}}(t) = \sum\limits_{m =  - \infty }^{ + \infty } {{j^{k + m}}} {d_k}(m){g_{\mathrm{C},k}}(t - m{T_k}/2){e^{j2\pi {f_{\mathrm{C},k}}t}},
		\label{eq:oqam}
	\end{equation}	
	where $d_{k}(m)$ is $m$-th real-valued data drawn from the real and imaginary parts of the mapped symbols which are independently and identically distributed (i.i.d.), $f_{\mathrm{C},k}$ is the central frequency, $T_{k}$ is the complex-valued symbol interval, and $ g_{\mathrm{C},k}(t) $ is a CE pulse-shaping filter with real and symmetric properties. 
	Let $T_{\mathrm{s}}$ be the sampling interval, and then the discrete-time version of \eqref{eq:oqam} is given by
	\begin{equation}
		{\chi}_{\mathrm{D},k}(n) = \sum\limits_{m =  - \infty }^{ + \infty } {{j^{k + m}}} {d_k}(m){g_{\mathrm{D},k}}(n - m{\Phi _k}/2){e^{j2\pi {f_k}n}},
		\label{eq:discrete_oqam}
	\end{equation} 
	where ${\Phi _k} = {T_k}/{T_{\mathrm{s}}}$ is the sampling factor, $f_{k}=T_{\mathrm{s}} f_{\mathrm{C},k}$ is the normalized central frequency,  ${\chi}_{\mathrm{D},k}(n)= {\chi}_{\mathrm{C},k}\left(n T_{\mathrm{s}}\right)$, and ${g_{\mathrm{D},k}}(n) = {g_{\mathrm{C},k}} \left( {n{T_{\mathrm{s}}}} \right)$. 
	
	In order to avoid the ISI caused by delay spread, we adopt the block-based transmission mechanism and insert the CP between successive transmission blocks \cite{Gao2011Cyclic, Wang2013CP}.
	We suppose $N_{\mathrm{c}}$ is the number of signal samples over each transmission block excluding the CP. 
	Accordingly, each transmission block of the $k$-th
	UE compromises $2N_{\mathrm{d},k}$ real-valued data symbols, where ${N_{\mathrm{d},k}} = {N_{\mathrm{c}}}/{\Phi _k}$. 
	Due to the CP insertion, the linear convolution in \eqref{eq:discrete_oqam} can be replaced by a circular one, then the transmitted signal of the $k$-th UE in the $l$-th block can be written as
	\begin{align}
		\label{eq:cyclic_oqam}
		&{{\chi}_{k,l}}(n) =\hspace{-2mm} \sum\limits_{m = 0}^{2{N_{\mathrm{d},k}} - 1} 
		\hspace{-2mm}
		{{j^{k + m}}} {e^{j\pi {f_k}m{\Phi _k}}}{d_{k,l}}(m)
		{g_k}\left( {{{\left \langle n - m{\Phi _k}/2 \right \rangle }_{{N_{\mathrm{c}}}}}} \right), \nonumber\\
		& \qquad\qquad\qquad\qquad\qquad\qquad\qquad\qquad 
		0 \le n\le {N_{\mathrm{c}}} - 1,
	\end{align}
	where $d_{k,l}(m) \triangleq d_{k}\left(2 l N_{\mathrm{d},k}+m\right)$ is the transmitted data of $k$-th UE in the $l$-th block, and $g_{k}(n)$ is defined as\cite{Wang2013CP} 
	\begin{equation}\label{eq:gkn}
		\resizebox{.50\textwidth}{!}{$
			g_k(n) = \left\{
			\begin{aligned}
				& g_{\mathrm{D},k}(n) e^{j 2\pi f_k n}, && 0 \leq n  \leq \frac{N_{\rm c}}{2} - 1, \\
				& g_{\mathrm{D},k}(n - N_{\mathrm{c}}) e^{j 2\pi f_k (n - N_{\mathrm{c}})}, && \frac{N_{\rm c}}{2} \leq n  \leq N_{\mathrm{c}} - 1.
			\end{aligned}
			\right.
			$}
	\end{equation} 

	To facilitate the vector expression of transmitted signals, we define the following vectors: 
	\begin{subequations}
		\begin{align}
			&\boldsymbol{\chi}_{k,l} \triangleq \left[{\chi}_{k,l}(0), {\chi}_{k,l}(1), \cdots , {\chi}_{k,l}\left(N_{\mathrm{c}}-1\right)\right]^{T},\\
			&\mathbf{g}_{k} \triangleq \left[g_{k}(0), g_{k}(1), \cdots , g_{k}\left(N_{\mathrm{c}}-1\right)\right]^{T},\\
			&\mathbf{d}_{k,l} \triangleq \left[d_{k,l}(0), d_{k,l}(1), \cdots , d_{k,l}\left(2 N_{\mathrm{d},k}-1\right)\right]^{T}. 
		\end{align}
	\end{subequations} 
	Based on the above definitions, the vector version of transmitted signal in \eqref{eq:cyclic_oqam} can be expressed as
	\begin{align}
		\label{eq:vector trans sig}
		\boldsymbol{\chi}_{k,l} = j^k \tilde{\mathbf{G}}_k \mathbf{M}_k \mathbf{\Omega}_k \mathbf{d}_{k,l},
	\end{align}
	where the circulant matrix $\tilde{\mathbf{G}}_k = \operatorname{circ}(\mathbf{g}_k)$ is generated from $\mathbf{g}_k$, the zero-padding interpolation matrix $ \mathbf{M}_k = [\mathbf{e}_{N_{\mathrm{c}},0}, \mathbf{e}_{N_{\mathrm{c}},\Phi_k/2}, \ldots, \mathbf{e}_{N_{\mathrm{c}},(2N_{{\rm d},k}-1)\Phi_k/2}] $, and the frequency modulation diagonal matrix $ \mathbf{\Omega}_k = \mathsf{diag}([1, e^{j\pi(f_k \Phi_k+0.5)}, \ldots, e^{j\pi(2N_{{\rm d},k}-1)(f_k \Phi_k+0.5)}]) $. 
	To facilitate the implementation and optimal design of the proposed scheme, we define the normalized center frequency as $f_k = \tfrac{a'_k - \lfloor (N_{{\rm d},k}+1)/2 \rfloor - \delta_k}{N_{\mathrm{c} }}$, where $a'_k$ is an integer index and $\delta_k$ equals $0.5$ and $0$ for even and odd $N_{{\rm d},k}$, respectively. 
	Therefore, $\mathbf{\Omega}_k$ can be further expressed as
	\begin{equation}
		\mathbf{\Omega}_k \hspace{-1mm}=\hspace{-1mm} \mathsf{diag}\Bigg(\Big[1, e^{j\pi \tfrac{a'_k}{N_{{\rm d},k}}}, \ldots, e^{j\pi \tfrac{(2N_{{\rm d},k}-1)a'_k}{N_{{\rm d},k}}}\Big]\Bigg)\cdot \boldsymbol{\Theta}_{2N_{{\rm d},k}},
	\end{equation}
	where $ \boldsymbol{\Theta}_{2N_{{\rm d},k}}$ is a $2N_{{\rm d},k} \times 2N_{{\rm d},k}$ diagonal matrix with its $n$-th diagonal entry given by $e^{-j\pi n /(2N_{{\rm d},k})}$.

	\subsection{Compatible Implementation Structure}\label{TCompatible} 
	
\begin{figure*}[pt]
		\centering
		\includegraphics[scale=0.6]{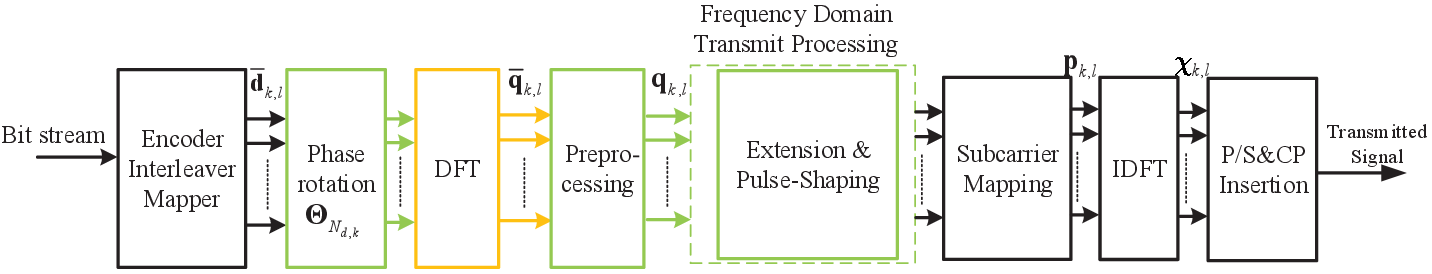}
		\caption{The compatible transmit structure of the proposed CE-CP-OFDMA within the framework of CP-OFDMA for the $k$-th UE. Black blocks: the basic CP-OFDMA. Black and yellow blocks: the DFT-s-OFDM based SC-FDMA. Black, yellow and green blocks: CE-CP-OFDMA.}
		\label{CPMSKOFDMkScdmasys2}
				\vspace{-5mm}
	\end{figure*}

	From the implementation structure of CE SC-FDMA signals in \eqref{eq:vector trans sig}, it is evident that its implementation structure is not directly compatible with the conventional CP-OFDMA system. Specifically, as shown in Fig.~\ref{CPMSKOFDMkScdmasys2}, the implementation structure of DFT-s-OFDM signals can be expressed as 
	\begin{align}
		\label{eq:ofdm trans sig}
		\boldsymbol{\chi}_{k,l}^{\rm dfts} = 
		\mathbf{W}_{N_{\rm c}}^{H}\mathbf{P}_{k}^{\rm sm}\mathbf{W}_{N_{{\rm d},k}}\bar{\mathbf{d}}_{k,l}, 
	\end{align}
	where ${{\bar{\bf{d}}}_{k,l}} = [{{\bar d}_{k,l}}(0),{{\bar d}_{k,l}}(1), \cdots ,{{\bar d}_{k,l}}({N_{\mathrm{d},k}} - 1)]^{T}$ denotes the transmitted complex-valued vector for the $l$-th data block of the $k$-th UE, with the $ m $-th symbol being defined as $ {{\bar d}_{k,l}}\left( m \right) = {d_{k,l}}(2m) + j{d_{k,l}}(2m + 1) $. 
	The subcarrier mapping matrix can be expressed by $ \mathbf{P}_{k}^{\rm sm} = \bar{\mathbf{\Lambda}}_{k}^{\rm sm} \mathbf{E}_{N_{{\rm d},k}}^{\rm sm} $, where $ \bar{\mathbf{\Lambda}}_{k}^{\rm sm} = \mathsf{diag}([\mathbf{0}_{1\times N_{1}},\mathbf{1}_{1\times N_{{\rm d},k}},\mathbf{0}_{1\times N_{2}}]^{T})  $ denotes the rectangular window frequency-domain filter matrix with $ N_{\rm c} = N_{1}+N_{2} + N_{{\rm d},k} $, and $ \mathbf{E}_{N_{{\rm d},k}}^{\rm sm} = \mathbf{1}_{\Phi_{k}\times 1}\otimes \mathbf{I}_{N_{{\rm d},k}} $ represents the frequency-domain periodic extension matrix. 
	Note that the DFT-s-OFDM waveform degrades to the conventional OFDM waveform when $ \mathbf{W}_{N_{{\rm d},k}} = \mathbf{I}_{N_{{\rm d},k}} $. 
	Considering the widespread utilization of OFDM waveform in existing wireless communication systems, this subsection explores the inherent characteristics of the proposed CE SC-FDMA waveform in~\eqref{eq:vector trans sig} and investigates its realization within the standard CP-OFDMA framework in line with \eqref{eq:ofdm trans sig}, i.e., CE-CP-OFDMA. 

	Considering the circulant characteristic of $ \tilde{\mathbf{G}}_{k} $ in \eqref{eq:vector trans sig}, we can operate eigenvalue decomposition as follows 
	\begin{align}
		\tilde{\mathbf{G}}_k = \mathbf{W}_{N_{\mathrm{c} }}^H \bar{\boldsymbol{\Lambda}}_k \mathbf{W}_{N_{\mathrm{c} }}, 
	\end{align}
	where the frequency-domain filter matrix $ \bar{\boldsymbol{\Lambda}}_k $ is a diagonal matrix with its diagonal vector being $ \bar{\boldsymbol{\lambda}}_k = \sqrt{N_{\mathrm{c} }} \mathbf{W}_{N_{\mathrm{c} }} \mathbf{g}_k $. By substituting the above equation into \eqref{eq:vector trans sig}, we can get 
	\begin{align}
		\label{eq:vector trans sig2}
		\boldsymbol{\chi}_{k,l} = j^k \mathbf{W}_{N_{\mathrm{c} }}^H \bar{\boldsymbol{\Lambda}}_k \mathbf{W}_{N_{\mathrm{c} }} 
		\mathbf{M}_k \mathbf{\Omega}_k \mathbf{d}_{k,l}. 
	\end{align}
	Subsequently, the frequency-domain signal $ \mathbf{W}_{N_{\mathrm{c} }} \mathbf{M}_k \mathbf{\Omega}_k \mathbf{d}_{k,l} $ can be interpreted as follows: the time-domain signal $ \mathbf{d}_{k,l} $ is first frequency-modulated by $ \mathbf{\Omega}_k $, then subjected to zero-padding interpolation via $ \mathbf{M}_k $, and finally transformed by DFT $ \mathbf{W}_{N_{\mathrm{c} }} $. Notably, the above operation is equivalent to performing a $2N_{{\rm d},k}$-point generalized DFT (GDFT) of the time-domain signal followed by a frequency-domain cyclic shift and periodic extension which can be expressed as 
	\begin{align}
		\label{eq:freq sig term}
		\mathbf{W}_{N_{\mathrm{c} }} \mathbf{M}_k \mathbf{\Omega}_k \mathbf{d}_{k,l} = 
		\sqrt{\frac{2}{\Phi_{k}}}
		{\mathbf{E}}_{2{N_{\mathrm{d},k}}}
		{\bf{S}}_k
		\tilde{\mathbf{W}}_{2 N_{\mathrm{d},k}} 
		\mathbf{d}_{k,l}, 
	\end{align} 
	where the frequency-domain cyclic shift matrix is defined as
	\begin{equation}
		{{\bf{S}}_k} = \left[ {\begin{array}{*{20}{c}}
			{\bf{0}}&{{{\bf{I}}_{{{ {\left \langle {{{a}'_k}} \right \rangle  } }_{2{N_{\mathrm{d},k}}}}}}}\\
			{{{\bf{I}}_{2{N_{\mathrm{d},k}} - {{{ {\left \langle {{{a}'_k}} \right \rangle  } }_{2{N_{\mathrm{d},k}}}}}}}}&{\bf{0}}
		\end{array}} \right],
	\end{equation}
	which is equivalent to the frequency modulation operation in the time domain. 
	${{\mathbf{E}}_{2{N_{\mathrm{d},k}}}} = {{\bf{1}}_{{\Phi _k}/2 \times 1}} \otimes {{\bf{I}}_{2{N_{\mathrm{d},k}}}}$ is the frequency-domain periodic extension matrix corresponding to the time-domain zero-padding interpolation. 
	$\tilde{\mathbf{W}}_{2 N_{\mathrm{d},k}}=\mathbf{W}_{ 2 N_{\mathrm{d},k}} \boldsymbol{\Theta}_{2 N_{\mathrm{d},k}}$ is the $2 N_{\mathrm{d},k}$-point generalized DFT (GDFT) matrix. Substituting \eqref{eq:freq sig term} into \eqref{eq:vector trans sig2}, we can obtain 
	\begin{align}
		\label{eq:vector trans sig3}
		\boldsymbol{\chi}_{k,l}=\mathbf{W}_{N_{\mathrm{c}}}^{H}  \mathbf{P}_{k} \mathbf{q}_{k,l},
	\end{align}
	where 
	\begin{subequations}
		\begin{align}
			\label{eq:P_k}
			&{\bf P}_k = {j^k}\sqrt {\frac{2}{{{\Phi _k}}}} {{\bar{\bf{\Lambda }}}_k}{{\mathbf{E}}_{2{N_{\mathrm{d},k}}}}{{\bf{S}}_k}, \\
			&\mathbf{q}_{k,l} = \tilde{\mathbf{W}}_{2 N_{\mathrm{d},k}} \mathbf{d}_{k,l}.
		\end{align}
	\end{subequations}
	Here, $ {\bf P}_k $ can be considered as the frequency-domain transmit processing (FDTP) matrix of the $k$-th UE. Note that both $ \mathbf{P}_{k}^{\rm sm} $ in \eqref{eq:ofdm trans sig} and $ {\bf P}_k $ in \eqref{eq:vector trans sig3} can be operated as linear signal processing with low complexity in the frequency domain.

	Comparing \eqref{eq:ofdm trans sig} and \eqref{eq:vector trans sig3}, it can be observed that both structures share a similar framework, differing only in the DFT operation. 
	Specifically, $ \mathbf{W}_{N_{{\rm d},k}}\bar{\mathbf{d}}_{k,l} $ in \eqref{eq:ofdm trans sig} can be viewed as the $ N_{{\rm d},k} $-point DFT for complex-valued symbols, while $ \mathbf{q}_{k,l} = \tilde{\mathbf{W}}_{2 N_{\mathrm{d},k}} \mathbf{d}_{k,l} $ in \eqref{eq:vector trans sig3} represents the $ 2N_{{\rm d},k} $-point GDFT for real-valued symbols. 
	In order to further improve the compatibility between the two structures, we aim to realize the computation of $ \mathbf{q}_{k,l} $ based on the $ N_{{\rm d},k} $-point DFT for complex-valued symbols with lower complexity. 
	Since $\mathbf{d}_{k,l}$ is real-valued, ${{\bf{q}}_{k,l}}$ exhibits the conjugate symmetry, which can be rewritten as 
	\begin{equation}\label{qul}
		{{\bf{q}}_{k,l}} = {[{({{\bf{s}}_{k,l}})^T}, {({{\bf{J}}_{{N_{\mathrm{d},k}}}}{\bf{s}}_{k,l}^*)^T}]^T},
	\end{equation}
	where $\mathbf{s}_{k,l} = [q_{k,l}(0), \ldots, q_{k,l}(N_{{\rm d},k}-1)]^T \in \mathbb{C}^{N_{{\rm d},k} \times 1}$ represents the effective information transmitted by the $k$-th UE in the $l$-th transmission block. 
	Based on \eqref{qul}, it is not difficult to derive the following relationship, 
	\begin{align}\label{sdfsf66}
		&{\bf{q}}_{k,l} = 
		\mathsf{Prep}\left( \bar{\mathbf{q}}_{k,l} 
		\right) \nonumber\\
		&=
		\left[
		\begin{array}{cc}
			{\bf{I}}_{N_{\mathrm{d},k}} & {\bf{I}}_{N_{\mathrm{d},k}} \\
			{\bf{I}}_{N_{\mathrm{d},k}} & -{\bf{I}}_{N_{\mathrm{d},k}} 
		\end{array}
		\right]
		\left[
		\begin{array}{cc}
			{\bf{I}}_{N_{\mathrm{d},k}} & {\bf{0}} \\
			{\bf{0}} & -j e^{-j\pi / (2N_{\mathrm{d},k})} {\bf{\Theta}}_{N_{\mathrm{d},k}} 
		\end{array}
		\right] \nonumber\\ 
		&\times \frac{1}{2\sqrt{2}}
		\left[
		\begin{array}{cc}
			{\bf{I}}_{N_{\mathrm{d},k}} & {\bf{J}}_{N_{\mathrm{d},k}} \\
			{\bf{I}}_{N_{\mathrm{d},k}} & -{\bf{J}}_{N_{\mathrm{d},k}} 
		\end{array}
		\right]
		\begin{bmatrix}
			\bar{\mathbf{q}}_{k,l} \\
			( \bar{\mathbf{q}}_{k,l} )^*
		\end{bmatrix},
	\end{align}
	where the operation $ \mathsf{Prep}(\cdot) $ can be viewed as preprocessing in the frequency domain and $ \bar{\mathbf{q}}_{k,l} = {{\bf{W}}_{{N_{\mathrm{d},k}}}}{{\bf{\Theta }}_{{N_{\mathrm{d},k}}}}{ {\bar{ \bf d}} _{k,l}} $. 

	Substituting \eqref{sdfsf66} into \eqref{eq:vector trans sig3}, we can finally obtain the compatible implementation structure of the proposed CE-CP-OFDMA scheme~as
	\begin{align}
		\label{eq:vector trans sig4}
		\boldsymbol{\chi}_{k,l}=\mathbf{W}_{N_{\mathrm{c}}}^{H}  \mathbf{P}_{k} \mathsf{Prep}\left( {{\bf{W}}_{{N_{\mathrm{d},k}}}}{{\bf{\Theta }}_{{N_{\mathrm{d},k}}}}{ {\bar{ \bf d}} _{k,l}} 
		\right). 
	\end{align} 
	The detailed implementation structure is illustrated in Fig.~\ref{CPMSKOFDMkScdmasys2}. 
	Specifically, for the transmitted signal of the $k$-th UE, the mapped symbol vectors ${{\bar{\bf{ d}}}_{k,l}}$ are first transformed into $ \bar{\mathbf{q}}_{k,l} $ through phase rotation and $N_{\mathrm{d},k}$-point DFT. After the operation of preprocessing, we can obtain ${{\bf{q}}_{k,l}} = \mathsf{Prep}(\bar{\mathbf{q}}_{k,l}) $ required in the proposed scheme. $\mathbf{q}_{k,l}$ is then multiplied by the FDTP matrix, yielding the frequency-domain transmitted signal vectors $ \mathbf{p}_{k,l} = \mathbf{P}_k\mathbf{q}_{k,l}$, including extension, pulse shaping, and subcarrier mapping. Finally, with the operation of the inverse DFT (IDFT),  the transmitted signal $ \boldsymbol{\chi}_{k,l} = \mathbf{W}_{N_{\mathrm{c}}}^{H}\mathbf{p}_{k,l} $ is generated. 
	
	{\itshape Remark 1:}
	The proposed CE-CP-OFDMA can be implemented based on the existing structures of the DFT-s-OFDM based SC-FDMA and CP-OFDMA with minimal modifications. Specifically, the proposed scheme incorporates additional low-complexity modules including phase rotation, preprocessing, extension, and pulse shaping compared to DFT-s-OFDM based SC-FDMA, while further introducing a DFT module relative to CP-OFDMA. 

	\subsection{Constant-Envelope Filter Optimization Design} 

	Motivated by the generation of CE MSK signals via half-sine pulse shaping in offset quadrature phase shift keying (QPSK) \cite{Gronemeyer}, we establish the following sufficient condition for CE modulation.
	
	\textit{Lemma 1:} When $d_{k}(m) \in \{ \pm 1\}$ and the pulse-shaping filter $ g_{\mathrm{C},k}(t) $ satisfies 
	\begin{equation} \label{eq:ce_filtering}
		\left\{ {\begin{array}{*{20}{l}}
				g_{\mathrm{C},k}^2(t) + g_{\mathrm{C},k}^2({T_k}/2 - t)  =  1, & \quad  0 \leq t <  \frac{T_k}{2}, \\
				g_{\mathrm{C},k}^2(t) = 0, & \quad  \left| t \right| \ge \frac{T_k}{2},
		\end{array}} \right.
	\end{equation} 
	the output signal $ {{\chi}_{\mathrm{C},k}}(t) $  is CE.
	Based on the condition in \eqref{eq:ce_filtering}, a special form of discrete function ${g_{\mathrm{D},k}}(t)$ for $k$-th UE can be written as 
	\begin{equation}
		\label{eq:ce_filtering2}
		{g_{\mathrm{D},k}}(n) = 
		\begin{cases}
			1, & n = 0, \\
			\cos \left( \theta_k(|n|) \right), & 1 \le |n| \le \frac{\Phi_k}{4} - 1, \\
			\frac{\sqrt{2}}{2}, & |n| = \frac{\Phi_k}{4}, \\
			\sin \left( \theta_k\left( \frac{\Phi_k}{2} - |n| \right) \right), & \frac{\Phi_k}{4} + 1 \le |n| \le \frac{\Phi_k}{2} - 1, \\
			0, & |n| \ge \frac{\Phi_k}{2},
		\end{cases}
	\end{equation}
	where ${{\theta _k}(n)}$ is the discrete-time phase parameter. 
	
	\textit{Remark 2:} Specifically, when  ${\theta _k}(n) = \frac{{\pi n}}{{{\Phi_k}}}$, ${g_{\mathrm{D},k}}(n)$ can be reduced to a half-sine function, i.e., 
	\begin{equation}
		{g_{\mathrm{D},k}}(n) = \left\{ {\begin{array}{*{20}{l}}
				{\cos \left( \frac{{\pi n}}{{{\Phi_k}}} \right),}&{ - \frac{{{\Phi_k}}}{2} < t < \frac{{{\Phi_k}}}{2}},\\
				{0,}&{{\rm{ otherwise}}{\rm{. }}}
		\end{array}} \right.
		\label{eq:half_sine_filtering}
	\end{equation}
	
	Different from OFDMA using band-limited frequency-domain filters to maintain inter-user orthogonality \cite{OFDMorW}, the equivalent frequency-domain pulse-shaping filter in \eqref{eq:ce_filtering2} is not strictly band-limited, potentially degrading orthogonality. To enable CE transmission while suppressing MAI, the pulse-shaping filter coefficients in the proposed CE-CP-OFDMA scheme should be optimized by minimizing stop-band energy. Exploiting the symmetry of \( g_{\mathrm{D},k}(n) \), the \( N_{\mathrm{c}} \)-point time-domain pulse-shaping filter coefficient vector is defined as follows 
	\begin{align}\label{eq:gpk}
		\mathbf{g}_{k} = [ 1, \mathbf{c}_{\rm filt}^{T}, \mathbf{0}_{1\times (N_{\rm c}-\Phi_{k}+1)}, ( \mathbf{J}_{\Phi_{k}/2-1} \mathbf{c}_{\rm filt} )^{T} ]^{T},
	\end{align}
	where $ \mathbf{c}_{\rm filt} = [ \cos \left( {{\theta _k}\left( 1 \right)} \right), \cdots ,\cos \left( {{\theta _k}\left( {{\Phi _k}/4 - 1} \right)} \right),\sqrt 2 /2 $,  $ \sin \left( {{\theta _k}\left( {{\Phi _k}/4 - 1} \right)} \right), \cdots ,\sin \left( {{\theta _k}\left( 1 \right)} \right) ]^{T} $. 
	The corresponding frequency-domain filter is given by
	\begin{equation}\label{eq:lambdapk}
		{\bar{\boldsymbol{\lambda}}_{k}} = \sqrt {{N_{\mathrm{c}}}} {{\bf{W}}_{{N_{\mathrm{c}}}}}{{\bf{g}}_{k}}\in \mathbb{C}^{N_{\mathrm{c}}\times 1}.
	\end{equation} 
	Since $\mathbf{g}_{k}$, as defined in \eqref{eq:gpk}, inherently ensures the CE property of the transmitted signal for the $k$-th UE, the phase parameters $ \boldsymbol{\theta}_{k} = [\theta_k(1), \dots, \theta_k(\Phi_k/4 - 1)]^{T}$ remain as free variables.  These degrees of freedom can be effectively exploited to minimize the stop-band energy of the filter, thereby reducing MAI without violating the CE constraint. 
	
	With the definition of $\left[ {\bar{N}_{{\rm d},k}/2,{N_{\mathrm{c}}}/2} \right]$  as the half stop-band range of the vector ${\bar{\boldsymbol{\lambda}}_{k}}$, the optimization problem is given by
	\begin{equation} \label{eq:min_leakage}
		\mathcal{Q}_1:\mathop {\min }\limits_{\boldsymbol{\theta}_{k}} \sum\limits_{i = \bar{N}_{{\rm d},k}/2 }^{{N_{\mathrm{c}}}/2} {{{\left| {\bar{\lambda}_{k}(i)} \right|}^2}}.
	\end{equation} 
	As \eqref{eq:min_leakage} is an optimization problem to obtain the minimum value of an unconstrained multi-variable function, the simplex method is one of the effective methods for such problems \cite{Jeffrey1998Convergence}. 
	The choice of $ \bar{N}_{{\rm d},k} $ makes a trade-off between the bandwidth of the transmit signals and the resulting out-of-band energy. 
	The frequency-domain transmitted signal exhibits the periodicity and conjugate symmetry. Specifically, the band of width $ N_{{\rm d},k} $ centered on the main lobe carries the signal vector $ \mathbf{s}_{k,l} $, the immediately neighboring band carries its conjugate symmetric component $ \mathbf{J}_{N_{{\rm d},k}}\mathbf{s}_{k,l}^{*} $, and subsequent outer band again carries $ \mathbf{s}_{k,l} $. Such alternating arrangement recurs periodically around the main-lobe center. 
	Considering the periodicity and conjugate symmetry of the transmitted signal, we choose to the preserve the main lobe of CE signal with $ \bar{N}_{{\rm d},k} = ( 2\bar{B} + 1 )N_{{\rm d},k} $, where $ \bar{B} $ is a non-negative integer. 
	
	To further reduce the out-of-band energy of the transmit signal, we further apply an additional Gaussian low-pass filter in the frequency domain $ \mathbf{b}_{k} \in \mathbb{C}^{N_{\rm c}\times 1} $ at the cost of slightly increasing the PAPR. Combing the CE filter with the Gaussian low-pass filter, the equivalent NCE frequency-domain filter can be expressed as 
	\begin{align}
		\label{eq:nce_filter}
		\mathbf{g}_{k}^{\rm nce} = \mathbf{g}_{k} \odot \mathbf{b}_{k}. 
	\end{align}
	The $ i $-th element of $ \mathbf{b}_{k} $ is defined as $ {b_{k}}(i) = {e^{ - \frac{{\ln 2}}{{8{{(\pi B_{\rm w})}^2}}}\omega _k^2(i)}} $, where 
	${\omega _k}(i) = 2\pi (\frac{i- a_k}{{{N_{\mathrm{c}}}}}  - \frac{1}{2})$ and  $B_{\rm w}$ is the corresponding $3$ dB bandwidth. 
	By adjusting $B_{\rm w}$, we have the freedom to make a trade-off between the PAPR and the signal bandwidth. 
	
\begin{figure}[htp]
		\centering
		\includegraphics[width=1.0\linewidth]{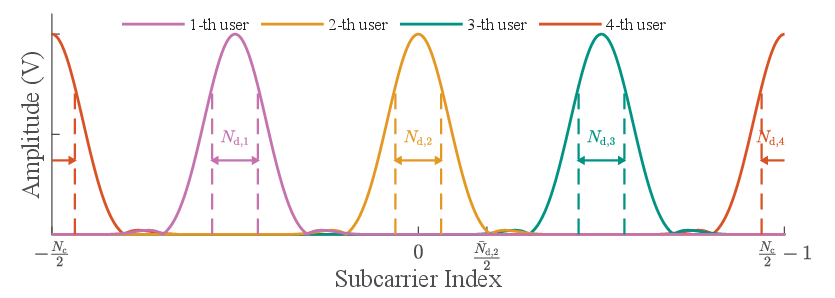}
		\caption{Frequency-domain subcarrier allocation for the proposed CE-CP-OFDMA transmission.}
		\label{fig:FDMA1}
	\end{figure}

	After the optimization of the filter coefficients in \eqref{eq:min_leakage} and \eqref{eq:nce_filter}, most energy of their transmit signals is concentrated on the limited bandwidth with $ \bar{N}_{{\rm d},k} $ subcarriers, which allows for a near orthogonal FDMA with the frequency-domain spacing of $ \bar{N}_{{\rm d},k} $ subcarriers for different UEs, as shown in Fig.~\ref{fig:FDMA1}. 
	As the CE signals are usually preferred in power-limited scenarios in practice, where the achievable signal-to-noise ratio (SNR) is relatively low, the performance degradation caused by such a non-orthogonal design is negligible, which will be validated in simulation results. 
	Note that the number of subcarriers allocated to the $k$-th UE, \(\bar{N}_{\mathrm{d},k}\),  exceeds the number of complex data symbols $N_{\mathrm{d},k} $, resulting in a loss of spectral efficiency compared to conventional OFDMA. However, in power-constrained scenarios such as NTN, the improved power efficiency achieved by the CE waveform outweighs the loss in spectral efficiency, which will also be validated in simulation~results.

	\section{ Receiver Design for CE-CP-OFDMA }\label{Section_4}
	In this section, we aim to design a receiver for the proposed CE-CP-OFDMA scheme over frequency-selective fading channels, which achieves high performance with low implementation complexity. 

	\subsection{Received Signal Model}\label{Received Signal Model}
	
	Let ${\bar{\mathbf{h}}_{k,l}} $ denote the frequency-domain channel vector of   the $k$-th UE in the $l$-th transmission block, 
	as defined in\cite{zhuang2025dmrs}
	\begin{equation}\label{hu}
		\bar{\mathbf{h}}_{k,l}=\sum_{q=0}^{Q_{k,l}-1} \alpha_{q,k,l} \mathbf{b}_{N_{\mathrm{c}}}\left(\tau_{q,k,l} \right)\in \mathbb{C}^{N_{\mathrm{c}}\times 1},
	\end{equation}
	where $Q_{k,l}$ denotes the number of path for the $k$-th UE in the $l$-th transmission block. $\alpha_{q,k,l}$ and $\tau_{q,k,l}$  represent the complex gain and the path delay of the $q$-th path for  the $k$-th UE in the $l$-th transmission block, respectively. The vector $\mathbf{b}_{N_{\mathrm{c}}}\left(\tau\right) \in \mathbb{C}^{N_{\mathrm{c}}\times 1}$ represents the delay-domain steering vector\cite{CiteSRSSCSI}, defined as
	\begin{gather}
		\mathbf{b}_{N_{\mathrm{c}}}\left(\tau\right)=\left[1, e^{-j 2 \pi \Delta f \tau}, \cdots, e^{-j 2 \pi\left(N_{\mathrm{c}}-1\right) \Delta f \tau}\right]^T,
	\end{gather}
	where $\Delta f$ represents the subcarrier spacing employed in the OFDM system.	We assume an uncorrelated fading environment, where	different propagation paths are independent\cite{xindaoalfa}. Therefore, the complex gain satisfies
	\begin{gather}
		\mathbb{E}\left\{ \alpha_{q,k,l}\alpha_{q',k,l}^* \right\}=\rho_{q,k,l}\delta\left[q - q'\right],
	\end{gather} 
	where $\rho_{q,k,l}$ denotes the power of the $q$-th path for the $k$-th UE in the $l$-th transmission block, and $\delta[\cdot]$ is the Kronecker delta function. 

\begin{figure*}[hpt]
		\centering
		\includegraphics[scale=0.65]{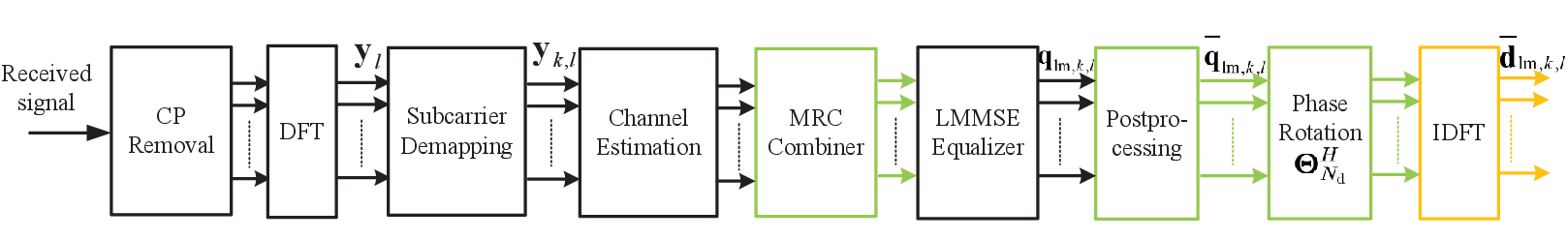}
		\caption{The receiver structure of the proposed CE-CP-OFDMA within the framework of CP-OFDMA. Black blocks: the basic CP-OFDMA. Black and yellow blocks: the DFT-s-OFDM based SC-FDMA. Black, yellow and green blocks: CE-CP-OFDMA.}
		\label{CPGMSKOFDM3receiver}
		\vspace{-10pt}
	\end{figure*}

	In the receiver as shown in  Fig.~\ref{CPGMSKOFDM3receiver}, based on \eqref{sdfsf66} and \eqref{eq:vector trans sig4}, the frequency-domain received signal in the  $l$-th block can be expressed as
	\begin{align}\label{equ:yl}
		\mathbf{y}_{l} &=\sum_{k=0}^{K-1} \bar{\mathbf{H}}_{k,l} \mathbf{P}_{k} \mathbf{q}_{k,l} + {{\bf{z}}_l}, 
	\end{align}
	where $ \bar{\mathbf{H}}_{k,l} = \mathsf{diag}\left( \bar{\mathbf{h}}_{k,l} \right) \in \mathbb{C}^{N_{\rm c}\times N_{\rm c}} $ represents the frequency-domain channel matrix of   the $k$-th UE in the $l$-th transmission block. ${{\bf{z}}_l}$ is an additive white Gaussian noise (AWGN) vector with zero mean and covariance $\sigma _z^2{{\bf{I}}_{{N_{\mathrm{c}}}}}$. 
	
	According to the frequency-domain subcarrier allocation depicted in Fig. \ref{fig:FDMA1}, the signals from different UEs are concentrated on the bandwidth with $ \bar{N}_{{\rm d},k} $ subcarriers and can be effectively separated with negligible MAI. 	
	Therefore, We adopt $\bar{N}_{\mathrm{d},k}$ subcarriers concentrated on the $k$-th UE's main lobe for equalization. 
	Considering the symmetry property of the CE filter, the subcarrier set occupied by the $k$-th UE can be written as 
	$\Gamma_k = \big\{ \langle a_k - \tfrac{\bar{N}_{\mathrm{d},k}}{2} + m \rangle_{N_{\mathrm{c}}} \;\big|\; m = 0, 1, \dots, \bar{N}_{\mathrm{d},k} - 1 \big\} $. 
	For convenience, we further define a one-to-one mapping $ \mathcal{M}_{k}(\cdot)$ from $\Gamma_k$ to the set $\{0, 1, \dots, \bar{N}_{\mathrm{d},k} - 1\}$, i.e.,
	$ \mathcal{M}_{k}( \langle a_k - \tfrac{\bar{N}_{\mathrm{d},k}}{2} + m \rangle_{N_{\mathrm{c}}} ) = m,$ 
	and $\mathcal{M}_{k}^{-1}(\cdot)$ denotes the inverse mapping.
	Combing \eqref{eq:P_k} and \eqref{qul}, the received signal vector for the $ k $-th UE can be further expressed as
	\begin{subequations}
		\begin{align}
			\label{equ:yul CE}
			&\text{Channel Estimation: }&&\mathbf{y}_{k,l} = \mathbf{X}_{k,l} {\boldsymbol{\Lambda}_{k}} \mathbf{h}_{k,l} + \mathbf{z}_{k,l}, \\
			\label{equ:yul EQU}
			&\text{Channel Equalization: } &&\mathbf{y}_{k,l} = \mathbf{H}_{k,l} {\boldsymbol{\Lambda}_{k}} \mathbf{x}_{k,l} + \mathbf{z}_{k,l},
		\end{align}
	\end{subequations} 
	where ${{\bf{y}}_{k,l}} \in\mathbb{C}^{\bar{N}_{\mathrm{d},k}\times 1}$ with its $m$-th element given by ${{{y}}_{k,l}}(m) = {{{y}}_l}\left( {\mathcal{M}_{k}^{ - 1}\left( m \right)} \right)$.  The corresponding channel matrix is defined as \( \mathbf{H}_{k,l} = \mathsf{diag}(\mathbf{h}_{k,l}) \), with \( \mathbf{h}_{k,l} \in \mathbb{C}^{\bar{N}_{\mathrm{d},k} \times 1} \) constructed by \( h_{k,l}(m) = \bar{h}_{k,l}(\mathcal{M}_{k}^{-1}(m)) \). Similarly, the diagonal filter matrix \( \boldsymbol{\Lambda}_k = \mathsf{diag}( \boldsymbol{\lambda}_{k} ) \in\mathbb{C}^{\bar{N}_{\mathrm{d},k}\times \bar{N}_{\mathrm{d},k}} \) with  \( \boldsymbol{\lambda}_{k}(m) = \bar{\lambda}_k(\mathcal{M}_{k}^{-1}(m)) \). 
	$ \mathbf{z}_{k,l} $ represents the corresponding noise vector. 
	Based on the periodicity and conjugate symmetry of the transmitted signal, $\mathbf{X}_{k,l} = \mathsf{diag}({{\bf{x}}_{k,l}}) $ and the $ m $-th $ N_{{\rm d},k} $-length subvector of $ \mathbf{x}_{k,l} $ is given by 
	\begin{align}
		\label{eq:xkl blk}
		\mathbf{x}_{k,l,m} = 
		\left\{
		\begin{array}{ll}
			\mathbf{s}_{k,l}, & \text{if}\: \langle m-\bar{B} \rangle_{2} = 0, \\
			\mathbf{J}_{N_{{\rm d},k}}\mathbf{s}_{k,l}^{*}, & \text{otherwise}.
		\end{array}
		\right.
	\end{align} 
	Taking $ \bar{B} = 1 $ as an example, $\mathbf{x}_{k,l} $ can be	expressed as $ \mathbf{x}_{k,l} = [ (\mathbf{J}_{N_{{\rm d},k}}\mathbf{s}_{k,l}^{*})^{T}, \mathbf{s}_{k,l}^{T} ,(\mathbf{J}_{N_{{\rm d},k}}\mathbf{s}_{k,l}^{*})^{T} ]^{T} $. 

	\subsection{Multi-Stage Channel Estimation}\label{channel estimation} 
	In this subsection, we propose a multi-stage channel estimation scheme for the proposed CE-CP-OFDMA system. Specifically, we firstly perform the low-complexity denoised channel estimation for the refined PDP parameter estimation, followed by the LMMSE channel estimation aided by the estimated PDP parameters. 
	Considering the negligible MAI, the discussion can be limited to the signal processing of a single UE. For simplicity, the subscript $k$ 
	and $l$ are omitted in the following analysis for simplicity. 
	Based on the received signal model for channel estimation \eqref{equ:yul CE}, the LMMSE channel estimation is given by 
	\begin{equation}\label{MMSE1}
		\mathbf{h}_{\rm lm} =\mathbf{R}_{\mathbf{h} } ( \mathbf{R}_{\mathbf{h}}+\sigma_{\rm z}^2 (\bar{\mathbf{X}}\bar{\mathbf{X}}^{H})^{-1} )^{-1} (\bar{\mathbf{X}})^{-1} \mathbf{y},
	\end{equation}
	where $ \bar{\mathbf{X}} = \boldsymbol{\Lambda}\mathbf{X} $ denotes the equivalent pilot matrix. the frequency-domain channel correlation matrix of the UE can be denoted by the PDP parameters  $ \{ {\rho}_q, {\tau}_q \}_{q=0}^{Q-1}  $ as 
	$ \mathbf{R}_{\mathbf{h} } 
	= \mathbb{E}\left\{ \mathbf{h} \mathbf{h}^H \right\} 
	= \sum_{q=0}^{Q-1} {\rho}_q \mathbf{b}_{\bar{N}_{\mathrm{d}}}({\tau}_q) \mathbf{b}_{\bar{N}_{\mathrm{d}}}^H({\tau}_q) $. 
	Under the given SNR, the LMMSE estimation accuracy is determined by the precision of channel and noise correlation~matrices.  

	In order to improve the accuracy of the PDP parameter estimation, we typically perform the low-complexity least squares (LS) channel estimation followed by DFT-based denoising as the input of PDP estimation module. 
	However, due to the non-faltness of the equivalent frequency-domain pilot matrix $ \boldsymbol{\Lambda} \mathbf{X} $, especially for the frequency-domain filter matrix $ \boldsymbol{\Lambda} $, the resulting noise in the DFT domain exhibits correlation, which limits the effectiveness of the DFT-based PDP estimation-aided MMSE channel estimation (DPMCE) to be introduced later. 
	To address this issue, we first compute the LS estimation of the equivalent channel $\boldsymbol{\xi} = \boldsymbol{\Lambda} \mathbf{h} $ as 
	\begin{align}
		\label{eq:ls_est}
		\boldsymbol{\xi}_{\rm ls} = ( \mathbf{X} )^{-1} \mathbf{y} = \boldsymbol{\xi} + \mathbf{X}^{-1} \mathbf{z}. 
	\end{align}
	We further address the non-faltness of the frequency-domain pilot matrix $\mathbf{X}$. To satisfy the CE requirement, the time-domain pilots in the propose CE-CP-OFDMA systems are constrained to be BPSK symbols. Therefore, the pilot design problem can be formulated as 
	\begin{align}
		\label{eq:Q2}
		&\mathcal{Q}_2: \min_{\mathbf{d}\in\mathbb{R}^{2N_{\rm d}\times 1}} 
		\Vert \mathbf{X}\mathbf{X}^{H} - \mathbf{I}_{\bar{N}_{\rm d}} \Vert_{\rm F}^{2}
		\nonumber\\
		&\quad \text{s.t.} \quad 
		d(m)  \in \left\{-1, 1\right\},  \forall m \in \{0, 1, \dots, 2N_{\rm d}-1\}. 
	\end{align}
	Considering the periodicity and conjugate symmetry of $ \mathbf{X} $, the minimization of $ \Vert \mathbf{X}\mathbf{X}^{H} - \mathbf{I}_{\bar{N}_{\rm d}} \Vert_{\rm F}^{2} $ is equivalent to the minimization of $ \sum_{i=0}^{N_{\rm d}-1}
	(| [\tilde{\mathbf{W}}_{2N_{\rm d}}]_{i,:} \mathbf{d}|^{2} - 1)^{2} $. 
	The optimization problem in \eqref{eq:Q2} is a typical binary optimization problem, which is generally NP-hard to solve optimally. Swarm-inspired algorithms, such as genetic algorithms and binary particle swarm optimization, are commonly employed to obtain approximate solutions~\cite{macedo2021overview}. 

	After the pilot design, since $\mathbf{G}$ is derived from a finite-tap time-domain filter, the equivalent channel ${\boldsymbol{\xi}}$ exhibits sparsity in the delay domain.
	Accordingly, we transform $\boldsymbol{\xi}_{\rm ls}$ to the delay domain and estimate the PDP as 
	\begin{align}\label{30}
		\mathbf{r}_{\tau} 
		\hspace{-1mm} = \hspace{-1mm}
		\mathsf{ReLU}
		(
		\mathbf{w}_\mathrm{rec} \odot   ( | 	\mathbf{W}_{\bar{N}_{\mathrm{d}}}^{H} 	{\boldsymbol{\xi }}_{\rm ls} |^{\odot 2} 
		\hspace{-1mm} - \hspace{-1mm}
		\sigma_{\rm z}^2 | \mathbf{W}_{\bar{N}_{\mathrm{d}}}^{H}  \mathbf{X}^{-1} \mathbf{1}_{\bar{N}_{\rm d}\times 1}  |^{\odot 2} )
		),
	\end{align}
	where $ \mathbf{w}_\mathrm{rec} \in \mathbb{R}^{\bar{N}_{\mathrm{d}} \times 1} $ is a rectangular window with ones corresponding to the CP duration~\cite{chuang}. 
	$ \mathsf{ReLU}(\cdot) $ is the rectified linear unit operator which clips the negative elements to zeros.

	For simplicity, under the assumption of uncorrelated equivalent channel taps in the DFT domain, the correlation matrix can be expressed as $ \mathbf{R}_{\tau} = \mathsf{diag} \left( \mathbf{r}_{\tau} \right) $, and the corresponding frequency-domain correlation matrix is given by $ \mathbf{R}_{\boldsymbol{\xi}} = \mathbf{W}_{\bar{N}_{\mathrm{d}}} \mathbf{R}_{\tau} \mathbf{W}_{\bar{N}_{\mathrm{d}}}^H $. 
	Substituting this into the LMMSE expression for the equivalent channel $\boldsymbol{\xi}$ yields 
	\begin{align}
		\label{MMSE2}
		\boldsymbol{\xi}_{\rm lm} 
		= \mathbf{W}_{\bar{N}_{\mathrm{d}}} \mathbf{R}_{\tau} ( \mathbf{R}_{\tau} + \sigma_{\rm z}^2 \mathbf{W}_{\bar{N}_{\mathrm{d}}}^H ( \mathbf{X} \mathbf{X}^H )^{-1} \mathbf{W}_{\bar{N}_{\mathrm{d}}} )^{-1}
		\mathbf{W}_{\bar{N}_{\mathrm{d}}}^H \boldsymbol{\xi}_{\rm ls}.
	\end{align} 
	Benefitting from the pilot design, the pilot matrix exhibits near orthogonality, i.e., $ \mathbf{X}\mathbf{X}^{H} \approx \mathbf{I}_{\bar{N}_{\rm d}} $, the LMMSE expression in \eqref{MMSE2} can be reduced to the low-complexity version as
	\begin{equation}\label{DFT2}
		\boldsymbol{\xi}_{\mathrm{dft}}=\mathbf{W}_{\bar{N}_{\mathrm{d}}}  \mathbf{R}_{\tau} \left( \mathbf{R}_{\tau} + \sigma_{\rm z}^2\mathbf{I}_{\bar{N}_{\mathrm{d}}}  \right)^{-1} \mathbf{W}_{\bar{N}_{\mathrm{d}}}^H \boldsymbol{\xi}_{\rm ls},  
	\end{equation}
	which is equivalent to the denoising in the DFT domain. 
	Accordingly, the DPMCE of $ \mathbf{h} $ is given by $ \mathbf{h}_{\rm dft} = (\boldsymbol{\Lambda})^{-1}\boldsymbol{\xi}_{\mathrm{dft}} $. 

	Consdiering the limited bandwidth, the resolusion of the DFT domain is insufficient and results in the energy leakage problem, which will degrades the channel estimation performance. 
	To address this issue, a more accurate estimation of the PDP parameters is presented in the following section via ESPRIT~\cite{ESPRIT}. 
	By leveraging the Vandermonde structure of $ \mathbf{h}_{\rm dft} $, the PDP parameters estimation problem can be formulated as
	\begin{align}
		\label{CPDecompositionProblem}
		\mathcal{Q}_3:	\min _{ \boldsymbol{\tau}, \boldsymbol{\alpha}_{\tau} }
		\Vert \mathbf{h}_{\rm dft} - \mathbf{B}_{\tau} \boldsymbol{\alpha}_{\tau} \Vert_{\rm F}^{2}, 
	\end{align}
	where 
	\begin{subequations}
		\begin{align}
			\label{eq:B tau}
			&\mathbf{B}_{\tau} = [ \mathbf{b}_{\bar{N}_{\rm d}}(\tau_{0}), \mathbf{b}_{\bar{N}_{\rm d}}(\tau_{1}), \ldots, \mathbf{b}_{\bar{N}_{\rm d}}(\tau_{Q-1}) ], \\
			&\boldsymbol{\tau} = [\tau_{0}, \tau_{1}, \ldots, \tau_{Q-1}]^{T}, \\
			&\boldsymbol{\alpha}_{\tau} = [\alpha_{\tau,0},\alpha_{\tau,1},\ldots,\alpha_{\tau,Q-1}]^{T}. 
		\end{align}
	\end{subequations} 
	Considering the transform-domain sparsity, the above problem is essentially a low-rank matrix decomposition problem compromising Vandermonde structure with generators $ \{ \phi_{q} = e^{-j2\pi\Delta f\tau_{q}} \}_{q=0}^{Q-1} $, which motivates using spatial-smoothing ESPRIT to efficiently solve the parameter estimation problem. 

	We firstly define the spatial smoothing parameters $ (K_{1},L_{1}) $ such that $ K_{1} + L_{1} = \bar{N}_{\rm d} + 1 $. Then, we apply these parameters to perform spatial smoothing\cite{CPD} on $ \mathbf{h}_{\rm dft} $, yielding 
	\begin{align}
		\label{eq:space smooth}
		\mathbf{H}_{\rm s} &= [ \tilde{\mathbf{J}}_{1}\mathbf{h}_{\rm dft}, 
		\tilde{\mathbf{J}}_{2}\mathbf{h}_{\rm dft}, 
		\ldots,
		\tilde{\mathbf{J}}_{L_{1}}\mathbf{h}_{\rm dft}] \nonumber\\
		&=\mathbf{B}_{\tau}^{(K_{1})} ( \mathbf{B}_{\tau}^{(L_{1})} \circ \boldsymbol{\alpha}_{\tau} )^{T} + \mathbf{N}_{\rm s}, 
	\end{align} 
	where $\tilde{\mathbf{J}}_{l_1} = [\mathbf{0}_{K_1 \times (l_1 - 1)}, \mathbf{I}_{K_1}, \mathbf{0}_{K_1 \times (L_1 - l_1)}] $ is the $ l_{1} $-th selection matrix~\cite{SelectionMatrix} and $ \mathbf{B}_{\tau}^{(K)} = [\mathbf{B}_{\tau}]_{0:K-1,:} $. $ \mathbf{N}_{\rm s} $ denotes the corresponding noise matrix. 

	Since the uniqueness condition is crucial for the decomposition problem in \eqref{CPDecompositionProblem}, we can derive the following uniqueness condition by leveraging Vandermonde structure~\cite{chen2024estimating} to enable spatial-smoothing ESPRIT as follows: 

	\textit{Lemma 2}: If the following conditions are satisfied: 1) $ \phi_{i} \neq \phi_{j},\: \forall i \neq j $ and 2) $ \mathsf{r}( \mathbf{B}_{\tau}^{(K_{1}-1)} ) = \mathsf{r} ( \mathbf{B}_{\tau}^{(L_{1})} \circ \boldsymbol{\alpha}_{\tau} ) = Q $, then the decomposition of $ \mathbf{H}_{\rm s} $ is unique. 
	Generically, the above conditions are equal to 
	\begin{align}
		\label{eq:VandermondeUniquenessCondition}
		\min \{ K_{1}-1,L_{1} \} \ge Q. 
	\end{align}
	The  uniqueness condition \eqref{eq:VandermondeUniquenessCondition} can be guaranteed by appropriately choosing parameters  $(K_1, L_1)$.

	Subsequently, a truncated singular value decomposition (SVD) is applied to $ \mathbf{H}_{\rm s} $, i.e.,
	\begin{align}
		\label{eq:SVD}
		\mathbf{H}_{\rm s}  = \mathbf{U} \boldsymbol{\Sigma} \mathbf{V}^{H}, 
	\end{align}
	where $\mathbf{U} \in \mathbb{C}^{K_1\times Q},  \boldsymbol{\Sigma} \in \mathbb{R}^{Q\times Q} $ and $ \mathbf{V} \in \mathbb{C}^{ L_1 \times Q} $ denote the truncated left singular vector matrix, singular value matrix, and right singular vector matrix, respectively. The value of $ Q $ is estimated based on the minimum description length (MDL) criterion~\cite{liu2016}. 
	Omitting the noise, according to \eqref{eq:VandermondeUniquenessCondition}, there exists a non-singular matrix $ \mathbf{M} \in \mathbb{C}^{{Q} \times {Q}} $ such that $ \mathbf{U}\mathbf{M} = \mathbf{B}_{\tau}^{(K_{1})} $ and $ \mathbf{V}^{*}\boldsymbol{\Sigma}(\mathbf{M}^{-1})^{T} = \mathbf{B}_{\tau}^{(L_{1})} \circ \boldsymbol{\alpha}_{\tau} $, which implies 
	\begin{align}
		\mathbf{U}_{1}\mathbf{M}\boldsymbol{\Phi} = \mathbf{U}_{2}\mathbf{M},
	\end{align} 
	where $ \mathbf{U}_{1} = [\mathbf{U}]_{0:K_{1}-2,:} $, 
	$ \mathbf{U}_{2} = [\mathbf{U}]_{1:K_{1}-1,:} $, and $ \boldsymbol{\Phi} = \mathsf{diag}( [\phi_{0},\phi_{1},\ldots,\phi_{Q-1}]^{T} ) $. 
	Considering the full-column rank of $ \mathbf{U}_{1}\mathbf{M} $, $ \mathbf{U}_{1} $ also has full column rank. 
	Therefore, we obtain 
	\begin{align}
		\mathbf{M}\boldsymbol{\Phi}\mathbf{M}^{-1} = \mathbf{U}_{1}^{\dagger}\mathbf{U}_{2}. 
	\end{align}
	By performing eigenvalue decomposition on  $\mathbf{U}_1^{\dagger} \mathbf{U}_2$, 
	we can get the normalized eigenvalues, i.e., the estimation of generators $ \{ \hat{\phi}_{q} \}_{q=0}^{Q-1} $, which can be further utilized for delay estimation as 
	\begin{align}
		\label{eq:tau est}
		\hat{\tau}_{q} = -\frac{\ln \hat{\phi}_{q}}{j2\pi\Delta f},\:\forall\: q=0,1,\ldots, Q-1. 
	\end{align}
	Then, the factor matrix $ \hat{\mathbf{B}}_{\tau} $ can be reconstructed based on \eqref{eq:B tau} and $ \boldsymbol{\alpha}_{\tau} $ can be estimated using LS criterion as 
	\begin{align}
		\label{eq:alpha est}
		\hat{\boldsymbol{\alpha}}_{\tau} = \hat{\mathbf{B}}_{\tau}^{\dagger}  \mathbf{h}_{\rm dft}. 
	\end{align}
	From \eqref{eq:tau est} and \eqref{eq:alpha est}, the frequency-domain correlation matrix can be reconstructed as 
	\begin{align}
		\label{Rh  PDP}
		\hat{\mathbf{R}}_{\mathbf{h}} = \hat{\mathbf{B}}_{\tau} \mathsf{diag}( \hat{\boldsymbol{\alpha}}_{\tau}^{\odot 2} ) \hat{\mathbf{B}}_{\tau}^{H}.
	\end{align}
	Substituting \eqref{Rh  PDP} into \eqref{MMSE1} yields the final LMMSE channel estimation. 

	It is worth noting that the LMMSE estimator in \eqref{MMSE1} requires high-dimensional matrix inversion, leading to high complexity. 
	To address this issue, we apply the Woodbury matrix identity and exploit the structure of the correlation matrix in \eqref{Rh PDP}, which yields 
	\begin{align}\label{MMSE}
		\hat{\mathbf{h}}_{\rm lm} 
		&=\hat{\mathbf{B}}_{\tau}
		( \hat{\mathbf{B}}_{\tau}^{H} \bar{\mathbf{X}}\bar{\mathbf{X}}^{H} \hat{\mathbf{B}}_{\tau} + \sigma_{\rm z}^{2}( \mathsf{diag}( \hat{\boldsymbol{\alpha}}_{\tau}^{\odot 2} ) )^{-1} )^{-1} 
		\hat{\mathbf{B}}_{\tau}^{H} \bar{\mathbf{X}}^{H} \mathbf{y}. 
	\end{align}
	Compared with \eqref{MMSE1}, \eqref{MMSE} reduces the matrix inversion complexity from $\mathcal{O}(\bar{N}_{\mathrm{d}}^3)$ to $\mathcal{O}(Q^3)$, which is particularly beneficial in sparse multipath scenarios (e.g., satellite communications). 
	The proposed ESPRIT-based PDP estimation-aided MMSE channel estimation (EPMCE) algorithm is summarized in Algorithm~\ref{TensorDecomposition-Based}. 
	
	\begin{algorithm}[!t]
		\caption{EPMCE Algorithm for the CE-CP-OFDMA}
		\label{TensorDecomposition-Based}
		\hspace*{0in} {\bf {Input:}} Pilot received signal $\mathbf{y} \in \mathbb{C}^{\bar{N}_d \times 1}$
		\begin{algorithmic}[1]
			\Statex \% DPMCE
			\State
			Compute  $ \mathbf{r}_{\tau}$ via \eqref{eq:ls_est}, \eqref{30}. 
			\State
			Compute  $ 	\boldsymbol{\xi}_{\rm dft} $ via \eqref{DFT2}.
			\State
			Compute $ \mathbf{h}_{\rm dft} = \boldsymbol{\Lambda}^{-1} \boldsymbol{\xi}_{\rm dft} $.
			\Statex \% PDP Parameter Estimation 
			\State  Compute the spatial-smoothing matrix $ \mathbf{H}_{\rm s}$  via \eqref{eq:space smooth}.
			\State  Compute the generators $ \hat{\boldsymbol{\tau}} $ via \eqref{eq:SVD}-\eqref{eq:tau est}.			
			\State
			Compute $ \hat{\boldsymbol{\alpha}}_{\tau} $ via \eqref{eq:alpha est}.
			\Statex \% LMMSE Channel Estimation
			\State
			Compute $ \hat{\mathbf{R}}_{\mathbf{h}} $ via \eqref{Rh  PDP}. 
			\State
			Compute $ \hat{\mathbf{h}}_{\rm lm} $ via \eqref{MMSE}.
		\end{algorithmic}
		\hspace*{0in} {\bf {Output:}} $	\hat{\mathbf{h}}_{\rm lm}\in \mathbb{C}^{\bar{N}_d \times 1}$.
	\end{algorithm}

	\subsection{Low-Complexity MRC-Aided LMMSE Equalization}
	Although the subvectors $ \{ \mathbf{x}_{m} \}_{m=0}^{2\bar{B}} $ of the transmitted signal $ \mathbf{x} $ carry the same data symbols for the CE transmission at the sacrifice of spectral efficiency, a MRC combiner can be designed to combine the corresponding subvectors $ \{ \mathbf{y}_{m} \}_{m=0}^{2\bar{B}} $ of the received signal $ \mathbf{y} $, thereby improving the received signal quality. 
	According to \eqref{equ:yul EQU} and \eqref{eq:xkl blk}, the $ m $-th received signal subvector can be further expressed as 
	\begin{align}
		\mathbf{y}_{m} = 
		\left\{
			\begin{array}{ll}
				\mathbf{A}_{m} \mathbf{s} + \mathbf{z}_{m}, &\text{if}\: \langle m-\bar{B} \rangle_{2} = 0, \\
				\mathbf{A}_{m} \mathbf{J}_{N_{\mathrm{d}}} \mathbf{s}^{*} + \mathbf{z}_{m}, &\text{otherwise},
			\end{array}
		\right.
	\end{align}
	where the diagonal matrix $ \mathbf{A}_{m} = \mathbf{H}_{m}\boldsymbol{\Lambda}_{m} $. Here, $ \mathbf{H}_{m} $ and $\boldsymbol{\Lambda}_{m}$ are both $N_{\rm d} \times N_{\rm d}$ diagonal matrices, whose diagonal elements are extracted from the $ mN_{\rm d} $-th to the $ ((m+1)N_{\rm d} - 1) $-th diagonal entries of $ \mathbf{H} $ and $ \boldsymbol{\Lambda} $, respectively. 
	Based on the structural properties of the received signal $ \mathbf{y}_{m} $, we firstly preprocess the received signal subvectors $ \mathbf{y}_{m} $ as follows: 
	\begin{align}
		\bar{\mathbf{r}}_{m} = 
		\left\{
			\begin{array}{ll}
				\mathbf{y}_{m}, &\text{if}\: \langle m-\bar{B} \rangle_{2} = 0, \\
				\mathbf{J}_{N_{\mathrm{d}}} \mathbf{y}_{m}^{*}, &\text{otherwise}.
			\end{array}
		\right.
	\end{align}
	Then, after the MRC, the resulting signal can be expressed as 
	\begin{align}
		\label{eq:MRC 1}
		\mathbf{r} = \sum_{m=0}^{2\bar{B}} \mathbf{W}_{m}^{\rm mrc} \bar{\mathbf{r}}_{m}
		= \sum_{m=0}^{2\bar{B}} \left( \mathbf{W}_{m}^{\rm mrc} \bar{\mathbf{A}}_{m}\mathbf{s} + \bar{\mathbf{z}}_{m} \right),
	\end{align}
	where $ \mathbf{W}_{m}^{\rm mrc} = \mathsf{diag}(\mathbf{w}_{m}^{\rm mrc}) \in \mathbb{C}^{N_{\rm d}\times N_{\rm d}} $ is the MRC matrix for the $ m $-th received signal subvector. 
	The diagonal matrix $ \bar{\mathbf{A}}_{m} = \mathbf{A}_{m} $ if $ \langle m-\bar{B} \rangle_{2} = 0 $, and $ \bar{\mathbf{A}}_{m} = \mathbf{A}_{m}^{*}  $ otherwise. 
	The diagonal vector of $ \bar{\mathbf{A}}_{m} $ is $ \bar{\mathbf{a}}_{m} $. 
	The noise term $ \bar{\mathbf{z}}_{m} = \mathbf{W}_{m}^{\rm mrc}\mathbf{z}_{m} $ if $ \langle m-\bar{B} \rangle_{2} = 0 $, and $ \bar{\mathbf{z}}_{m} = \mathbf{W}_{m}^{\rm mrc}\mathbf{J}_{N_{\mathrm{d}}} \mathbf{z}_{m}^{*} $ otherwise. 

	The next step is to design the combining matrix $ \{ \mathbf{W}_{m}^{\rm mrc} \}_{m=0}^{2\bar{B}} $ such  that the SNR of each element in $ \mathbf{r} $ is maximized. The corresponding optimization problem for $ r(i) $ can be formulated as
	\begin{align}
		\label{eq:max_w}
		\mathcal{Q}_{4}: \max_{\{\mathbf{w}_{m}^{\rm mrc}(i)\}_{m=0}^{2\bar{B}}}\: \mathsf{SNR}_{i},
	\end{align}
	where the SNR of $ r_{i} $ is given by 
	\begin{align}
		\mathsf{SNR}_{i} = \frac{ \sum_{m=0}^{2\bar{B}} | \mathbf{w}_{m}^{\rm mrc}(i) \bar{\mathbf{a}}_{m}(i) |^{2}  }{ \sum_{m=0}^{2\bar{B}} |\mathbf{w}_{m}^{\rm mrc}(i)|^{2} }
		\cdot \frac{\mathcal{E}_{\rm s}}{\sigma_{\rm z}^2},
	\end{align}
	where $\mathcal{E}_{\rm s} $ denotes the symbol power.  
	According to the Cauchy-Schwarz inequality, the possible optimal solution to \eqref{eq:max_w} is $ \mathbf{W}_{m}^{\rm mrc} = \bar{\mathbf{A}}_{m}^{*},\:\forall\: m $. 
	Substituting this result into \eqref{eq:MRC 1} leads to the following expression for the received signal after MRC
	\begin{align}
		\label{eq:MRC 2}
		\mathbf{r} = \mathbf{B}\mathbf{s} + \bar{\mathbf{z}},
	\end{align}
	where $ \mathbf{B} = \sum_{m=0}^{2\bar{B}} \bar{\mathbf{A}}_{m} \bar{\mathbf{A}}_{m}^{*} $, and $ \bar{\mathbf{z}} = \sum_{m=0}^{2\bar{B}} \bar{\mathbf{z}}_{m} $. 

	In order to effectively mitigate ISI, we adopt the LMMSE equalizer based on the received signal model in \eqref{eq:MRC 2}. Considering the i.i.d. characteristic of the transmitted data vector $ \mathbf{d} $, the LMMSE estimate of $ s(i) $ is expressed as 
	\begin{align}\label{sulmhat}
		s_{\rm lm}(i) &= [\mathsf{Cov}(\mathbf{s})]_{i,i} \mathbf{e}_{N_{\rm d},i}^{T}\mathbf{B}^{H}
		( \mathbf{B} \mathsf{Cov}(\mathbf{s}) \mathbf{B}^{H} + \sigma_{\rm z}^{2} \mathbf{B} )^{-1} \nonumber\\
		&\quad \times ( \mathbf{r} - \mathbf{B}\mathsf{E}(\mathbf{r}) ) + \mathsf{E}(s(i)). 
	\end{align}
	With the assumption that each element of $\mathbf{d}$ has zero mean and unit variance, \eqref{sulmhat} further simplifies to 
	\begin{align}\label{detect1}
		s_{\rm lm}(i) = \rho_{i}s(i) + \hat{z}_{i},
	\end{align}
	where 
	\begin{subequations}
		\begin{align}
			\rho_{i} &\hspace{-0.5mm}=\hspace{-0.5mm} \mathbf{e}_{N_{\rm d},i}^{T}\mathbf{B}^{H} 
			( \mathbf{B} \mathbf{B}^{H} \hspace{-0.5mm}+\hspace{-0.5mm} \sigma_{\rm z}^{2} \mathbf{B} )^{-1} 
			\mathbf{B} \mathbf{e}_{N_{\rm d},i}, \\
			\hat{z}_{i} &\hspace{-0.5mm}=\hspace{-0.5mm} \mathbf{e}_{N_{\rm d},i}^{T}\mathbf{B}^{H} 
			( \mathbf{B} \mathbf{B}^{H} \hspace{-0.5mm}+\hspace{-0.5mm} \sigma_{\rm z}^{2} \mathbf{B} )^{-1} 
			( \mathbf{B}( \mathbf{s} \hspace{-0.5mm}-\hspace{-0.5mm} s(i)\mathbf{e}_{N_{\rm d},i} ) \hspace{-0.5mm}+\hspace{-0.5mm} \bar{\mathbf{z}} ). 
		\end{align}
	\end{subequations}
	In \eqref{detect1}, the first term contains information about $s(i) $, while the second term represents interference and noise whose variance is $\rho_i(1-\rho_i)$. 
	After deriving the estimation $ \mathbf{s}_{\rm lm} $, the estimation $ \mathbf{q}_{\rm lm} $ can be  calculated based on \eqref{qul}. 
	It is evident that the inversion operations in the proposed equalizer are only performed on diagonal matrices. Consequently, the LMMSE equalizer illustrated in Fig.~\ref{CPGMSKOFDM3receiver} features low computational complexity and practical feasibility. 

	After equalization, the receiver performs similar processing operations as those used by the transmitter in \eqref{sdfsf66} to recover the transmitted symbol $ \bar{\mathbf{d}}_{\rm lm} $, which is given by
	\begin{align}
		\bar{\mathbf{d}}_{\rm lm} = \frac{1}{\rho_{m}} \mathbf{W}_{N_{\rm d}}^{H} \boldsymbol{\Theta}_{N_{\rm d}}^{H} \bar{\mathbf{q}}_{\rm lm},
	\end{align}
	where 
	\begin{align}
		\bar{\mathbf{q}}_{\rm lm} &= \frac{1}{2\sqrt{2}}
		\left[ {\begin{array}{*{20}{c}}
			{{{\bf{I}}_{{N_{\mathrm{d}}}}}}&{j{e^{j\pi /(2{N_{\mathrm{d}}})}}{\bf{\Theta }}_{{N_{\mathrm{d}}}}^*}
		\end{array}} \right]
		\left[ {\begin{array}{*{20}{c}}
			{{{\bf{I}}_{{N_{\mathrm{d}}}}}}&{{{\bf{I}}_{{N_{\mathrm{d}}}}}}\\
			{{{\bf{I}}_{{N_{\mathrm{d}}}}}}&{ - {{\bf{I}}_{{N_{\mathrm{d}}}}}}
		\end{array}} \right] \nonumber\\
		&\times \left( \mathbf{q}_{\rm lm} + \mathbf{J}_{2N_{\rm d}}(\mathbf{q}_{\rm lm})^{*} \right). 
	\end{align}
	After the subsequent steps of soft demodulation and decoding, the information bits can be recovered. 

	\textit{Remark 3}: Overall, Fig. \ref{CPGMSKOFDM3receiver} illustrates the compatible receiver structure of the proposed approach within the framework of CP-OFDMA. Specifically, the received signal
	first undergoes CP removal and ${{N_{\mathrm{c}}}}$-point DFT operation, followed by subcarrier demapping to obtain the frequency-domain received signal 
	$\mathbf{y}_{k,l}$ for the $k$-th UE. Channel estimation, MRC, and LMMSE equalization are then applied to obtain the equalized real-valued signal $ \mathbf{q}_{{\rm lm},k,l} $. This signal is postprocessed to reconstruct the complex-valued vector $ \bar{\mathbf{q}}_{{\rm lm},k,l} $. Finally, phase rotation with  ${\bf{\Theta }}_{{N_{\mathrm{d}}}}^H$ and $N_{\mathrm{d},k}$-point IDFT are applied to recover the transmitted symbol vector $ \bar{\mathbf{d}}_{{\rm lm},k,l} $. 
	The proposed CP-CE-OFDMA receiver can be implemented based on existing OFDM receiver structures, requiring only the replacement of the conventional frequency-domain processing module with the proposed low-complexity processing scheme. 
	
	\section{ Downlink Multiple Access Transmission Scheme Based on Constant-Envelope Waveform}\label{Section_5} 

	Existing research on CE waveforms has largely focused on the uplink. However, for power-limited systems such as NTN, suppressing the downlink PAPR is equally critical. Notably, without specific downlink design, FDMA induces time-domain superposition of user data streams, thereby increasing PAPR. To overcome this limitation, we aim to develop novel downlink multiple access transmission schemes based on the proposed CE waveform. 

	\subsection{Symbol-Level and Bit-Level Multiple Access} 

\begin{figure*}[hpt]
		\centering
		\includegraphics[scale=0.65]{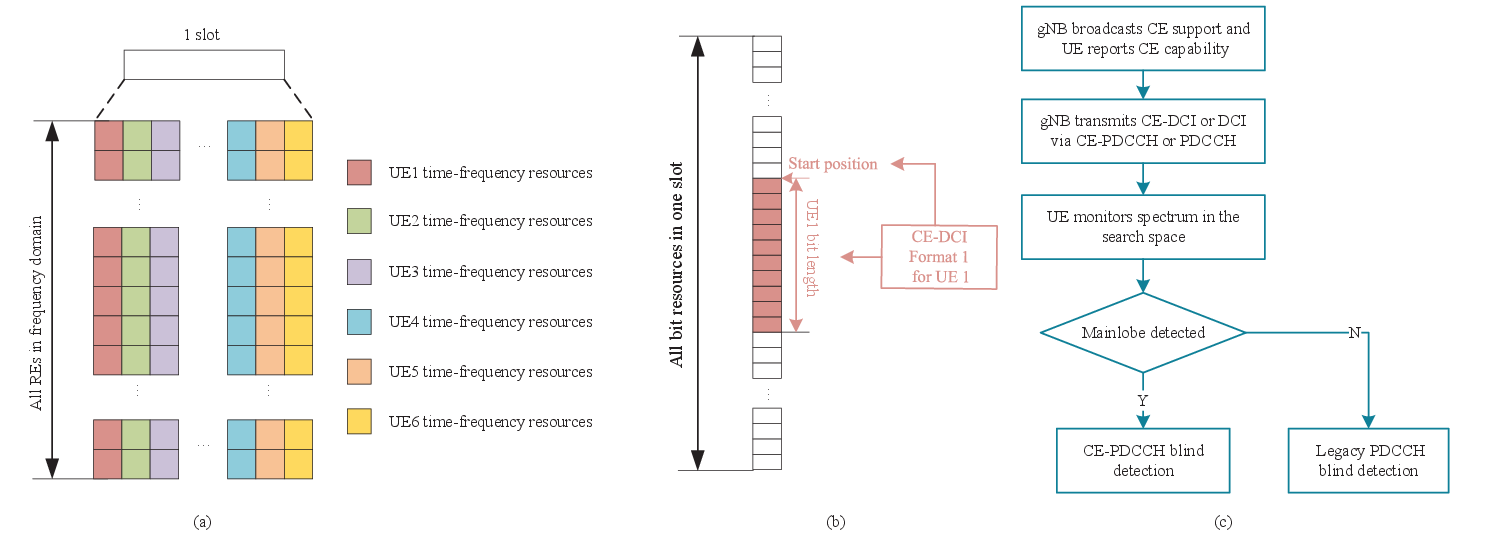}
		\caption{(a) Resource allocation in the SLMA scheme. (b) Resource indication method of the BLMA scheme. (c) Procedure of the proposed system-level downlink CE transmission scheme.}
		\label{fig:ce-transmission}
		\vspace{-5pt}
	\end{figure*}

	The SLMA scheme maps data of different UEs to distinct OFDM symbols, thereby avoiding direct superposition of multi-user signals. The corresponding resource allocation is illustrated in Fig.~\ref{fig:ce-transmission}(a). 
	Symbol-level resource allocation primarily concerns pilot and data signals, while the configuration of control signals remains consistent with the NR standard and will be discussed in detail in Section~\ref{Sys Design}.
	To reduce the pilot overhead of CE, the pilots are transmitted  only in slots with scheduled CE control or data signals. Specifically, the pilots are allocated to the first OFDM symbol, enabling the UE to efficiently process subsequent CE control or data signals within the same slot. 
	CE data allocations strictly avoid overlap with common reference signals in NR. 

	The BLMA scheme arranges the encoded bits of different UEs according to a predefined  rule,  which are then jointly mapped to the transmission resources. 
	Specifically, the high layer schedules slot-wise transmission resources for the downlink multiple access transmission, with the first OFDM symbol for pilots and the remaining symbols for data. 
	The transmission resources are then partitioned into bit segments, with each UE's encoded bits mapped to distinct segments.  
	To achieve scheduling flexibility comparable to that of the NR system, the proposed bit scheduling scheme takes the number of bits carried by one resource block in as the basic scheduling unit. Due to the QPSK modulation constraint for the proposed CE waveform, each bit segment carries $24$ bits. 
	Provided a slot carrying $N$ data bits, the data transmission resources can be divided into $\left \lfloor \frac{N}{24} \right \rfloor $ bit segments. 
	Similar to the bandwidth part (BWP) configuration, the higher layer can configure up to four bit segment intervals (BSIs) for each UE, where each BSI is defined by a start index and a length, as shown in Fig.~\ref{fig:ce-transmission}(b). 

	\subsection{Constant-Envelope DCI Design} 
	After the design of downlink multiple access mechanism for the proposed CE waveform, we need to design the dedicated CE-DCI format 0 and format 1 to enable the SLMA and BLMA, respectively. The detailed DCI designs are as follows. 

	Based on the mapping structure of SLMA, the SLMA processing flow consists of three phases: 1) The base station (BS) assigns multi-user data to distinct OFDM symbols; 2) The BS transmits CE-DCI format 0 carrying user data resource allocation indicators; 3) Upon successful blind detection of CE-DCI format 0, the corresponding symbol position is identified and CE reception is executed for data recovery. The proposed CE-DCI format 0 is constructed through field-level adaptations of the DCI format 1\_0 with cell radio network temporary identifier (C-RNTI) scrambling, as specified in third generation partnership project (3GPP) Release 17~\cite{38212}. 
	Key modifications include:

	\begin{itemize}
		\item 
		{Frequency-Domain Resource Assignment}: Symbol-level multiplexing enables the receiver to directly acquire  $\bar{N}_{\mathrm{d}} $ subcarriers from the BWP start point, removing frequency-domain indications.
		\item 
		{Time-Domain Resource Assignment}: This field specifies slot offset $k$, consecutive slots $L$, and OFDM symbol index $j$. The UE applies $k$-slot offset after DCI reception, then receives the $j$-th OFDM symbol in $L$ consecutive slots.  The parameter bitwidth is configured through radio resource control (RRC) signaling based on predefined allocation tables.
		\item {Modulation and Coding Scheme}: Since the proposed CE waveform supports only QPSK modulation, this field is repurposed to indicate the code rate, requiring only $3$ bits.
	\end{itemize}
	The proposed design reduces DCI overhead by removing the frequency-domain resource assignment field and compressing the modulation and coding scheme field, at the expense of scheduling flexibility. 

	Building on this bit-level mapping structure, the BLMA processing involves three  phases: 
	1) The BS assigns multi-user data to distinct bit segments corresponding to the slot-wise transmission resource; 
	2) The BS transmits CE-DCI format 1 specifying the start position and length of the bit segment allocated to the UE. 
	3) Upon successful blind detection of CE-DCI format 1, UE performs CE reception at indicated bit-segment locations. 
	Similarly, the proposed CE-DCI format 1 is designed through field-level modifications of the DCI format 1\_0. Key modifications include:
	\begin{itemize}
		\item 
		{Frequency-Domain Resource Assignment}: Consistent with the SLMA,  this field can be removed.
		\item 
		{Time-Domain Resource Assignment}: Consistent with legacy NR system, this field occupying $2$ bits indicates the slot offset ($0,1,2,3$) for CE transmission scheduling.
		\item 
		{Bit-Domain Resource Assignment}: This is a newly added field, used to indicate the location of the UE's bit information, specifying the starting index and length of the bit segment. The field size is calculated as $\left\lceil\log_2(N_\mathrm{seg}(N_\mathrm{seg}+1)/2)\right\rceil$,  where $ N_\mathrm{seg}$ is the length of the BSI assigned to the UE.
		\item 	{Modulation and Coding Scheme}: Consistent with the SLMA, this field occupies $3$ bits. 
	\end{itemize} 
	The BLMA scheme exhibits marginally increased control overhead compared to SLMA, while achieving enhanced resource scheduling flexibility. 
	Similar to the calculation scheme and size configuration of NR's frequency-domain assignment, the proposed bit-domain resource assignment field introduces no additional signaling overhead.

\subsection{System-Level Design of Constant-Envelope Transmission}\label{Sys Design} 

To introduce new transmission modes without significantly increasing UE reception complexity while  maintaining  compatibility with the NR system, we propose a dedicated CE-PDCCH for CE-DCI transmission. 
This design decouples the blind detection of CE-DCI from that of legacy NR DCI, thereby enabling new DCI formats to be introduced without increasing the complexity of legacy blind detection procedures. 	
To further ensure compatibility with the NR system, the CE-PDCCH shall conform to existing NR-defined control resource set (CORESET) and search space configurations. 
Note that since each BWP can support up to three CORESETs~\cite{38331}, at least one CORESET must span more than $\bar{N}_{\mathrm{d}}$ subcarriers to satisfy CE reception requirements for CE-supported UEs. 

Given that CE-PDCCH shares CORESET and search space configurations with legacy PDCCH, it is critical for the UE to promptly identify CE-PDCCH within the search space. As illustrated in Fig.\ref{fig:ce-transmission}(c), the receiver differentiates CE-PDCCH from legacy PDCCH within the search space by analyzing the spectral characteristics of the received signals. Specifically, a signal exhibiting a distinct mainlobe pattern, as shown in Fig.\ref{fig:FDMA1}, corresponds to CE-PDCCH and triggers CE-PDCCH blind detection, while a flat spectral profile corresponds to legacy PDCCH and triggers standard blind detection. 

Building upon the CE-PDCCH compatibility design, we propose a system-level downlink CE transmission scheme. As illustrated in Fig.~\ref{fig:ce-transmission}(c), the implementation process is described as follows 
\begin{enumerate}
	\item The BS broadcasts whether it supports the CE transmission mode. If CE mode is supported, the UE indicates its CE reception capability following RRC connection establishment.
	\item  If the BS supports CE mode, the BS transmits CE-DCI through the CE-PDCCH to indicate the CE transmission mode (SLMA or BLMA scheme) and the resource location of the UE data in the CE physical downlink shared channel (CE-PDSCH).
	\item For CE-supported UEs, spectral monitoring in the PDCCH search space is performed prior to blind detection to determine whether to activate CE reception mode. 
	\item  Under NR legacy mode, the UE receives legacy DCI to schedule legacy PDSCH. In the CE reception mode, the UE relies on CE-DCI for CE-PDSCH scheduling, with CE-DCI defined in two formats that mirror the two multiple-access schemes.
\end{enumerate}
To conclude, the proposed CE transmission scheme achieves low-complexity implementation without compromising compatibility with the existing NR system.

\section{Simulation Results}\label{Section_6}

This section compares the spectrum  characteristics and evaluates the envelope fluctuation and bit error rate (BER) performance of the proposed CE-CP-OFDMA with the conventional OFDMA and DFT-s-OFDM based SC-FDMA. 
In the simulation, we assume uniform transmission parameters across all UEs. 
Unless otherwise specified, the basic simulation parameters are presented in Table \ref{simuTable:Parameters}.

\begin{table}[t]
	\footnotesize
	\centering
	\setlength{\tabcolsep}{6pt} %
	\caption{Simulation Setup Parameters}
	\label{simuTable:Parameters}
	\begin{tabular*}{\linewidth}{@{\extracolsep{\fill}}ll} %
		\toprule
		Parameter  &  Value \\
		\midrule
		Channel model & AWGN or NTN-TDL-D~\cite{38811} \\
		LDPC code rate & $379/1024$ \\
		Modulation scheme & QPSK \\
		Maximum number of UEs  & $K=4$ \\
		Number of complex symbols per block & $N_{\mathrm{d}}=256$ \\
		Number of subcarriers & $N_{\mathrm{c}}=4096$ \\
		Subcarrier spacing $\Delta f$ & 120 kHz\\
		Roll-off factor for DFT-s-OFDM & $\gamma=0,0.25,1$ \\
		Factor of the Gaussian low-pass filter & $B_{\rm w}\cdot T_{k}=1$ \\
		\bottomrule
	\end{tabular*}
\end{table}

\subsection{Spectrum Characteristic and PAPR Performance}

\begin{figure}[t]
	\centering
	\includegraphics[width=3in]{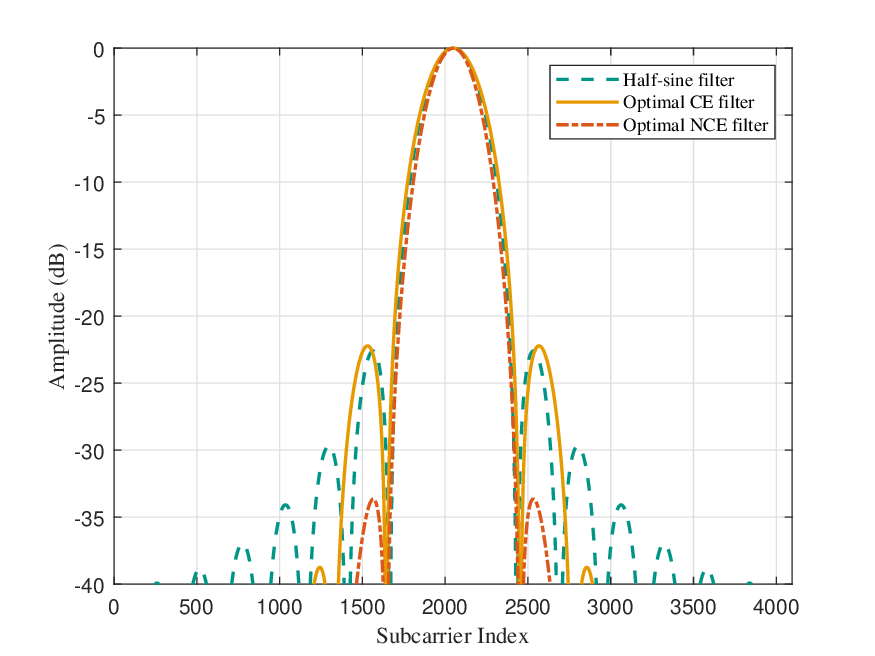}
	\caption{The amplitude-frequency spectrum of the half-sine filter, the optimal CE filter, and the optimal NCE filter. }
	\label{simufig:Compare_CENCE}
\end{figure}

Fig.~\ref{simufig:Compare_CENCE} illustrates the amplitude-frequency spectrum of several pulse-shaping filters. The proposed optimal CE filter exhibits significantly reduced sidelobes compared to the conventional half-sine filter, thereby enhancing spectral confinement.
To further suppress out-of-band leakage, a Gaussian low-pass filter is cascaded after the CE filter, forming a composite filter referred to as the optimal NCE filter. 
This configuration achieves a sidelobe attenuation of $-33.6$ dB, which satisfies the in-band emission requirement specified in 3GPP technical specification (TS) 38.101-2~\cite{38101}. Consequently, the signal shaped by the NCE filter can be effectively confined within a bandwidth $ \bar{N}_{\rm d} = 3N_{\rm d} $ following the NR specification, which are utilized in the subsequent simulation. 

\begin{figure}[t]
	\centering
	\includegraphics[width=3in]{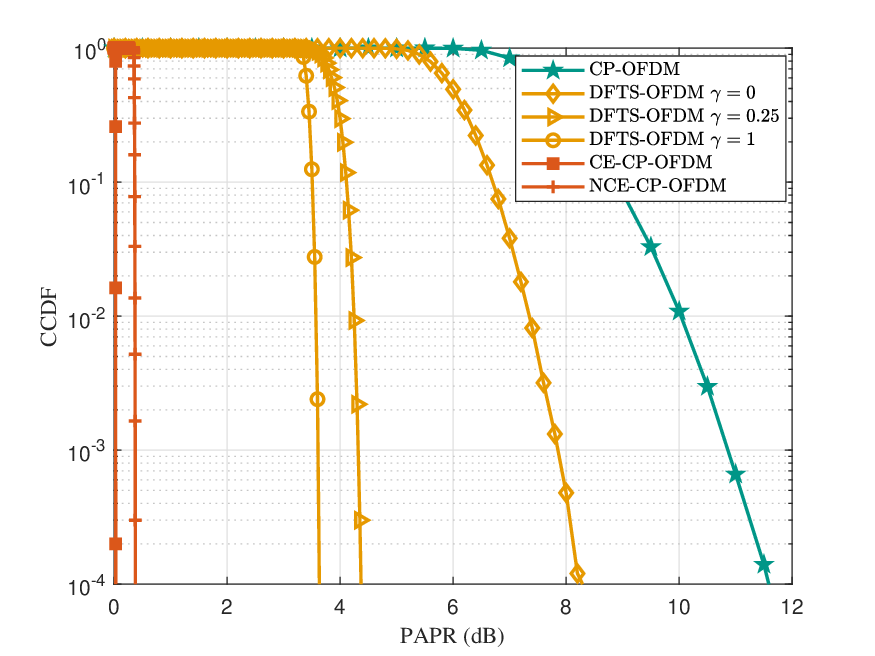}
	\caption{PAPR performance comparison of different waveforms.}
	\label{simufig:PAPR_Compare}
\end{figure}

Fig.~\ref{simufig:PAPR_Compare} illustrates the PAPR performance of the four waveforms. It can be clearly observed that the proposed schemes yield significant reduction in PAPR. 
Considering $ \text{CCDF} = 10^{-3} $ as a reference, the CE-CP-OFDM waveform achieves an ultra-low PAPR of $ 0$ dB due to its CE property. 
Although slightly relaxing the CE constraint, the NCE-CP-OFDM waveform largely reduces the out-of-band energy while still achieving a low PAPR of only $0.37$ dB. 
Compared with the DFT-s-OFDM waveform,  the PAPR of NCE-CP-OFDM is reduced by $7.5$ dB, $4.0$ dB, and $3.3$ dB for roll-off factors of $0$, $0.25$, and $1$, respectively. Greater gains are observed over the conventional CP-OFDM waveform, with a PAPR reduction of $10.7$ dB achieved by the proposed NCE-CP-OFDM.

\subsection{Channel Estimation Performance}\label{fs} 

\begin{figure}[t]
	\centering
	\includegraphics[width=3in]{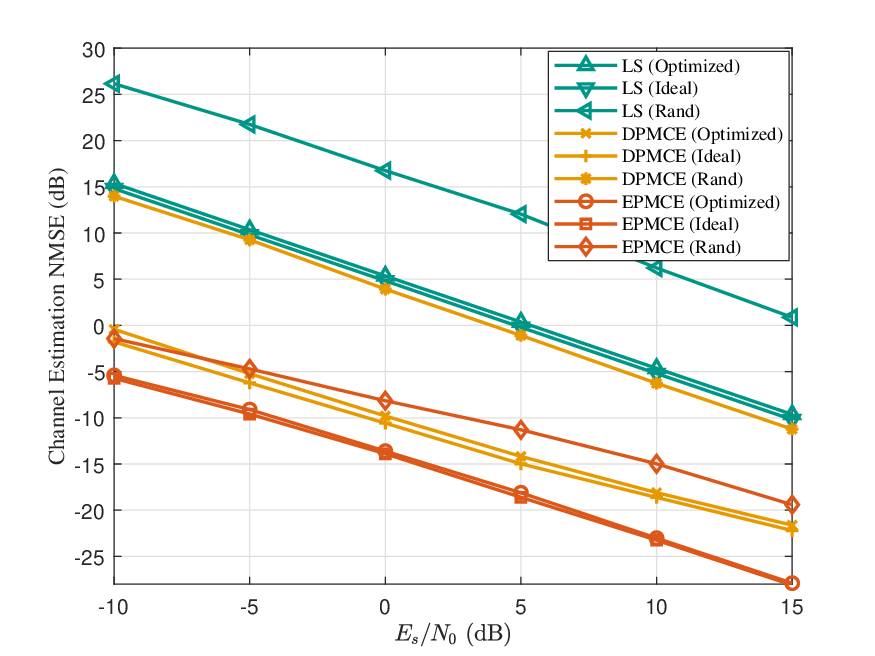}
	\caption{Channel estimation performance with different pilots and  methods.}
	\label{CH}
\end{figure}

For channel estimation performance evaluation, we adopt the NTN tapped delay line model with delay profile D (NTL-TDL-D) channel model defined in 3GPP technical report 38.811~\cite{38811}, which is a representative frequency-selective fading channel for NTN scenarios. According to Table 6.7.2-1b of the report, the delay spread is typically set to $37$ ns to reflect propagation conditions in NTN environments. 
To illustrate the performance of channel estimation, we first define the normalized mean square error (NMSE) performance metric 
\begin{equation}
	\text{NMSE} = 10 \log_{10} \left( \frac{\| \hat{\mathbf{h}} - \mathbf{h} \|_{2}^{2}}{\| \mathbf{h} \|_{2}^{2}} \right),
\end{equation}
where $\bf{h} $ and $\hat{\bf{h}} $ are the true and estimated frequency-domain channel response vector, respectively. 

To evaluate the performance of the optimized pilot sequence, we compare it against the following two baseline pilot~schemes:
\begin{itemize}
	\item \textbf{Ideal}: The ideal pilot sequence is spectrally flat but without the time-domain CE property. Given the importance of spectral flatness to channel estimation, it serves as the theoretical performance ceiling under the same estimator.  
	\item \textbf{Rand}: The random pilot sequence is randomly generated time-domain binary sequences with CE property. Without any optimization, its frequency-domain amplitude is generally non-flat, serving as the lower bound for CE~pilots. 
\end{itemize}

Fig.~\ref{CH} presents a comprehensive comparison of channel estimation performance using three methods: LS, DPMCE, and EPMCE in Algorithm \ref{TensorDecomposition-Based}, under different pilot schemes. It can be observed that the substantial performance gaps exist between the random and ideal pilots. Specifically, at $E_{\rm s}/N_0 = 0$ dB, the corresponding performance gaps are approximately $13$ dB, $13$ dB, and $5.5$ dB for LS, DPMCE, and EPMCE, respectively, which is consistent with the theoretical analysis discussed in Section \ref{channel estimation}. 
Moreover, the estimation performance achieved with the optimized pilot closely approaches that of the ideal pilot across all methods, validating the effectiveness of the proposed optimized pilot design. 
In addition, when all three methods operate under the optimized pilot and an $E_{\rm s}/N_0$ of $0$ dB, EPMCE yields an $18$ dB NMSE improvement over LS, primarily due to LS's sensitivity to noise amplification in the low-gain regions of the filter~$\mathbf{G}$. Compared to DPMCE, EPMCE achieves an additional $4$ dB gain, as it avoids energy leakage during PDP estimation, leading to more accurate channel reconstruction. 

\subsection{BER Performance}
\begin{figure}[t]
	\centering
	\includegraphics[width=3in]{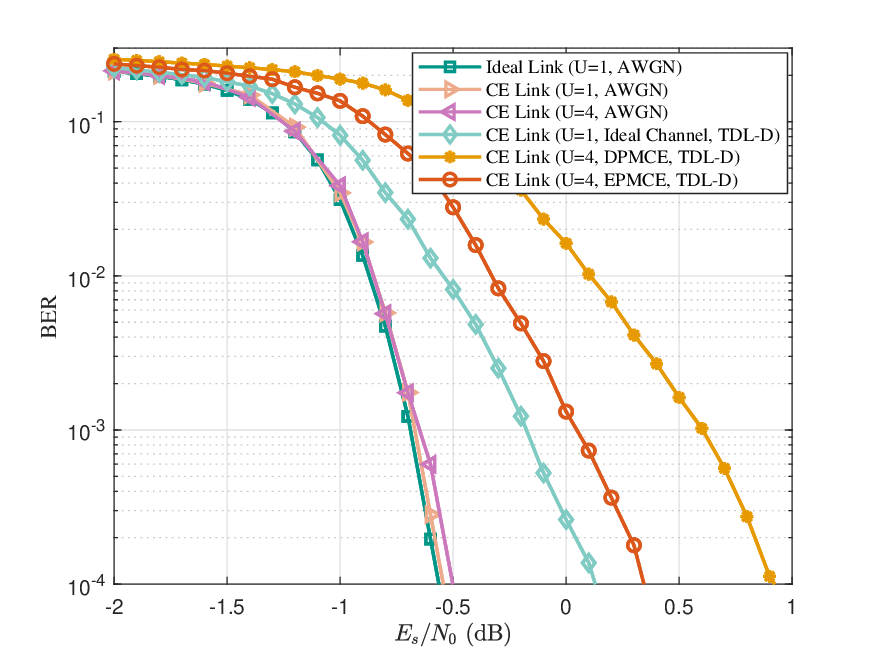}
	\caption{BER performance of the proposed CE scheme under AWGN and NTN-TDL-D channels.}
	\label{BER1}
\end{figure}

Fig.~\ref{BER1} shows the BER performance of the proposed CE-CP-OFDM scheme  under AWGN and NTN-TDL-D channels. In the figure, 	\lq U=1\rq\ and \lq U=4\rq\  indicate systems with one and four UEs, respectively, while \lq CE Link\rq\ and \lq Ideal Link\rq\ represent the proposed CE transmission link and the ideal QPSK-modulated OFDM link. 
In the AWGN channel, for the single-user case, the CE link achieves nearly identical BER to the ideal link  after MRC combining, confirming negligible performance loss. 
Furthermore, BER performance in the multi-user scenarios remains comparable to the single-user case, which confirms that the proper main-lobe spacing $ \bar{N}_{\rm d} = 3N_{\rm d} $ enables multi-user CE uplink transmission with negligible MAI. 
Under the NTN-TDL-D channel, the BER performance with EPMCE approaches that with perfect channel knowledge within $0.2$ dB and reduces the required $E_{\rm s}/N_0$ by approximately $0.7$ dB at a BER of $10^{-4}$, highlighting the robustness of the proposed CE transmission scheme in NTN scenarios. 

\begin{figure}[t]
	\centering
	\includegraphics[width=3in]{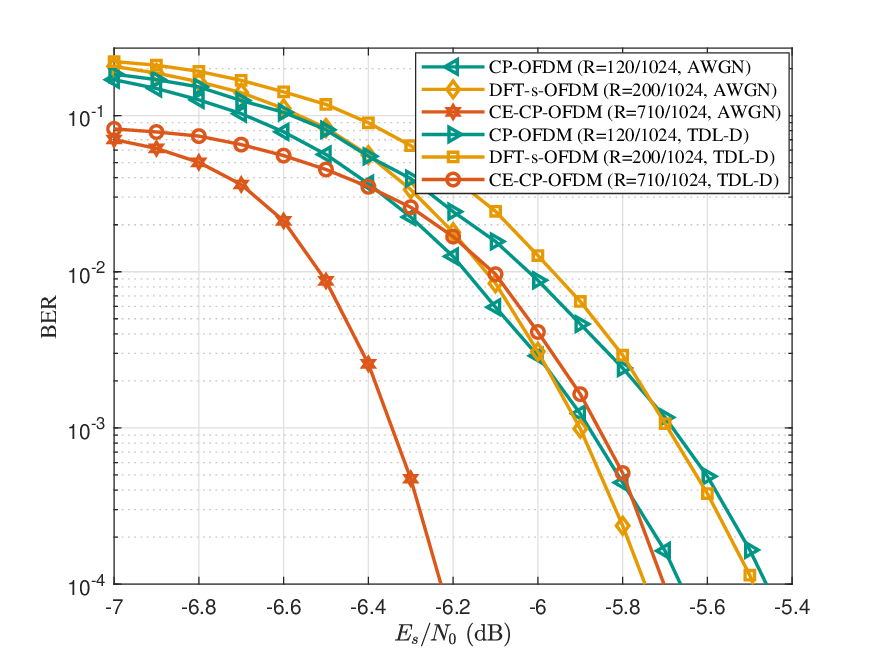}
	\caption{BER performance comparison of CE-CP-OFDM, DFT-s-OFDM ($\gamma=0$), and CP-OFDM under power back-off with equal transmit power and bandwidth.}
	\label{BER2}	
\end{figure}

Fig.~\ref{BER2} illustrates the BER performance of CE-CP-OFDM, DFT-s-OFDM ($\gamma = 0$), and CP-OFDM under power back-off conditions, where all waveforms are transmitted with equal power and bandwidth over both AWGN and NTN-TDL-D channels. 
The power back-off levels are determined based on the power reduction specified in 3GPP TS 38.101-2~\cite{38101}, with CE-CP-OFDM offering $3$ dB and $5$ dB transmission power advantages over DFT-s-OFDM and CP-OFDM, respectively. 
At the target BER of $10^{-4}$, CE-CP-OFDM achieves bit rates of approximately 2 and 1.2 times those of CP-OFDM and DFT-s-OFDM, respectively, in consideration of the spectral sacrifice caused by the CE modulation. 
Furthermore, relative to the baseline waveforms, the required demodulation threshold $E_{\rm s}/N_0$ is reduced by $ 0.5 $ dB and $ 0.2 $ dB under the AWGN and NTN-TDL-D channels, respectively.  
These results demonstrate that CE-CP-OFDM offers significant advantages in power-limited scenarios such as NTN. 

\section{Conclusion}\label{Section_Conclusion6}
In this paper, we proposed a CE transmission scheme compatible with the CP-OFDMA framework. Firstly, we developed a CE waveform implementation that maintains CP-OFDMA compatibility, and optimized the pulse-shaping filter under the CE constraint to suppress out-of-band energy. Subsequently, we proposed the multi-stage channel estimation with optimized pilot design to effectively address the channel estimation challenges introduced by non-flat frequency-domain pilot and filter. 
By exploiting the periodicity and conjugate symmetry of CE signals, we designed an MRC combiner followed by a low-complexity frequency-domain LMMSE equalizer. 
Based on the proposed CE waveform, we further designed a multi-user downlink transmission scheme that complies with the NR standard.
Numerical results demonstrate that the proposed scheme achieves near-ideal BER performance while maintaining low transceiver complexity.

\bibliographystyle{IEEEtran}
\bibliography{CKMPaperRef}

\newpage

\end{document}